\documentclass[fleqn,notitlepage,12pt]{article}

\pdfoutput=1

\usepackage{amsmath,amsfonts,amssymb,graphicx,psfrag}
\usepackage[colorlinks=true,urlcolor=blue,citecolor=black,linkcolor=blue]{hyperref}
\usepackage{setspace} 
\usepackage{dcolumn}
\newcolumntype{.}{D{.}{.}{-1}}
\newcolumntype{d}[1]{D{.}{.}{#1}}
\usepackage{natbib}
\bibpunct{(}{)}{;}{a}{}{,}
\usepackage{comment}
\usepackage{multirow}
\usepackage{lscape}
\usepackage{bbm}
\usepackage{scalefnt}
\usepackage[autolanguage]{numprint}
\usepackage{ctable}
\usepackage{paralist}
\usepackage{appendix}
\usepackage{multirow}

\usepackage{booktabs}
\usepackage{bm}
\usepackage{threeparttable}
\usepackage[top=1in, left=1in, right=1in,
bottom=1in]{geometry}

\usepackage{lmodern}
\fontfamily{lmtt}\selectfont
\usepackage[T1]{fontenc}

\usepackage{rotating}
\usepackage{subfig}
\usepackage{etoolbox}

\makeatletter

\patchcmd{\NAT@test}{\else \NAT@nm}{\else \NAT@nmfmt{\NAT@nm}}{}{}

\DeclareRobustCommand\citepos
  {\begingroup
   \let\NAT@nmfmt\NAT@posfmt
   \NAT@swafalse\let\NAT@ctype\z@\NAT@partrue
   \@ifstar{\NAT@fulltrue\NAT@citetp}{\NAT@fullfalse\NAT@citetp}}

\let\NAT@orig@nmfmt\NAT@nmfmt
\def\NAT@posfmt#1{\NAT@orig@nmfmt{#1's}}
\renewcommand{\P}{{\text{Pr}}}
\newcommand{\E}{{{E}}}

\newcommand{\bZ}{{\mathbf{Z}}}
\newcommand{\bA}{{\mathbf{A}}}
\newcommand{\bS}{{\mathbf{S}}}
\newcommand{\bB}{{\mathbf{B}}}

\newcommand{\br}{{\mathbf{r}}}

\newcommand{\sign}{{\text{sign}}}

\newcommand{\bz}{{\mathbf{z}}}

\newcommand{\bY}{{\mathbf{Y}}}

\newcommand{\1}{\mathbbm{1}}

\newcommand{\bR}{{{R}}}

\newcommand{\cF}{{\mathcal{F}}}
\newcommand{\cC}{{\mathcal{C}}}
\newcommand{\cR}{{\mathcal{R}}}

\newcommand{\cZ}{{\mathcal{Z}}}

\newcommand{\bq}{\mathbf{q}}

\newcommand{\indep}{\perp\!\!\!\perp}

\makeatletter
\renewcommand{\section}{\@startsection{section}{1}{0em}{\baselineskip}{0.5\baselineskip}{\large\bfseries\large}}
\renewcommand{\subsection}{\@startsection{subsection}{0}{0em}{\baselineskip}{0.5\baselineskip}{\normalfont\bfseries\normalsize}}
\makeatletter

\newcolumntype{.}{D{.}{.}{-1}}
\newcolumntype{d}[1]{D{.}{.}{#1}}

\begin{document}
\pagestyle{plain}

\singlespacing
\newcommand{\blind}{0}

\newcommand{\tit}{Patterns of Effects and Sensitivity Analysis for Differences-in-Differences\thanks{The authors thank Paul Rosenbaum for useful conservations on this topic and many others.}}

\if0\blind

{\title{\tit}

\author{Luke J. Keele\thanks{Associate Professor, University of Pennsylvania,
            3400 Spruce St.,
            Philadelphia, PA 19104 Email: luke.keele@gmail.com, corresponding author.}
\and Dylan S. Small\thanks{Professor, University of Pennsylvania, 400 Huntsman Hall, 3730 Walnut St., Philadelphia, PA 19104.  E-mail: dsmall@wharton.upenn.edu }     
\and Jesse Y. Hsu\thanks{Assistant Professor, University of Pennsylvania, 423 Guardian Dr, Philadelphia, PA 19104, Email: hsu9@mail.med.upenn.edu}      
\and Colin B. Fogarty\thanks{Assistant Professor, Massachusetts Institute of Technology, Cambridge, MA 02142, Email: cfogarty@mit.edu}
}

\date{\today}

\maketitle
}\fi

\if1\blind
\title{\bf \tit}
\maketitle
\fi

\maketitle

\thispagestyle{empty}

\begin{abstract}
Applied analysts often use the differences-in-differences (DID) method to estimate the causal effect of policy interventions with observational data. The method is widely used, as the required before and after comparison of a treated and control group is commonly encountered in practice. DID removes bias from unobserved time-invariant confounders. While DID removes bias from time-invariant confounders, bias from time-varying confounders may be present. Hence, like any observational comparison, DID studies remain susceptible to bias from hidden confounders. Here, we develop a method of sensitivity analysis that allows investigators to quantify the amount of bias necessary to change a study's conclusions. Our method operates within a matched design that removes bias from observed baseline covariates. We develop methods for both binary and continuous outcomes. We then apply our methods to two different empirical examples from the social sciences. In the first application, we study the effect of changes to disability payments in Germany. In the second, we re-examine whether election day registration increased turnout in Wisconsin.
\end{abstract}

\begin{center}
\end{center}

\clearpage
\doublespacing
\section{Assessing the causal effects of policy changes with examples from health policy and election administration}

Effective policymaking requires understanding the causal effects of proposals in order to devise the optimal policy. In almost every policy domain, including health, labor, education, environmental studies, and public safety, data and statistical methods are used to estimate the causal effects of policy interventions. Health policy and election administration are two specific areas of public policy where investigators have sought to understand the effects of specific policy changes.

For example, many countries have sick pay and disability insurance programs.  These programs provide some level of pay to workers when they miss work due to illness or disability. Policymakers face trade-offs when determining the generosity of disability insurance programs. One concern is that if these programs are too generous, workers will miss work for non-illness related reasons. In Germany, disability benefits once covered 100\% of wages for the first six weeks of sickness. In 1995, Germany changed employment regulations such that disability payments for all workers were reduced from 100\% coverage to 80\% coverage. Workers who were covered by a collective bargaining contract (unionized workers) disputed the change in the courts and were excluded from the changes. \citet{puhani2010} exploited the fact that unionized workers were exempt from the change to understand whether the less generous payments led to workers using disability services at lower rates.

In the U.S., state and local officials set many of the regulations related to the administration of elections including the voter registration process. It is generally assumed that when states make the voter registration process more onerous, citizens are less likely to vote. Thus state policy may have important effects on the level of democratic participation. For example, some states allow voters to register to vote on election day, while most states require voter registration to be completed at least 2-4 weeks before election day. A large literature has sought to estimate whether election day voter registration (EDR) leads to higher levels of turnout \citep{Brians:1999,Brians:2001,Hanmer:2007,Hanmer:2009,Highton:1998,Knack:2001, Mitchell:1995, Rhine:1995, Teixeira:1992, Timpone:1998, Wolfinger:1980, leighley2013votes}. Both of these applications represent areas of public policy where future changes in policy depend on understanding the extent to which policy affects behavior. As we outline next, studies of policy change often rely on a common statistical device known as differences-in-differences (DID) to estimate causal effects.

\subsection{Assessing the effects of policy changes using differences-in-differences to control for unobserved time invariant confounders}

It is well understood that randomized policy evaluations are the ``gold-standard,'' since randomization ensures that subjects are similar except for receipt of the treatment of interest. However, many policy evaluations occur in settings where randomized experiments are difficult or impossible. For example, states have made many changes to voter registration systems and these have not ever, to our knowledge, been evaluated with a randomized trial. The primary alternative to randomized trials are observational studies where the objective remains to elucidate cause-and-effect relationships in contexts but where subjects select their own treatment status \citep{Cochran:1965}. 

Of course, when treatments are not randomized, differences in outcomes may reflect pretreatment differences in treated and control groups rather than treatment effects \citep{Cochran:1965,Rubin:1974}. Pretreatment differences arise for either measurable reasons which form overt biases or unmeasured reasons which may cause hidden bias. In an observational study, analysts use pretreatment covariates and a statistical adjustment strategy such as matching or regression modeling to remove overt biases in the hopes of consistently estimating treatment effects. Unfortunately, such statistical adjustments do little to ensure that estimated treatment effects do not reflect hidden bias from confounders that were not included in the statistical adjustments. As such, investigators often employ devices consisting of information collected in hopes of distinguishing an estimated association from bias \citep{Rosenbaum:2010}.  One such device is the method of differences-in-differences (DID) which can be applied when the investigator collects data on the treated and control groups before and after the treatment is administered. The DID estimate of the treatment effect is the difference between the after-minus-before responses for the treated group and the after-minus-before responses for the control group.  The advantage of DID is that--subject to a set of assumptions--bias from both observed and \emph{unobserved} time-invariant confounders is removed from the treatment effect estimate.

DID is widely used to study the effects of policy changes. In both of the applications outlined above, DID was applied to estimate the effect of the policy change of interest. There is a particularly long history of DID in the study of policy effects. One of the earliest uses of DID was to estimate the effect of a minimum wage law in Oregon that led to higher wages in Portland but not the rest of the state \citep{obenauer1915effect}. One famous example based on DID studied the effect of the Mariel Boatlift from Cuba on employment rates in the Miami labor market \citep{card1990impact}. Another well known example based on differences-in-differences is in \citet{Dynarski:1999}. Here, she studies the treatment effect of the additional aid on the decision to attend college, using changes in the Social Security Student Benefit Program, which awarded college aid to high school seniors with deceased fathers of Social Security recipients. Other examples include \citepos{Card:1994} study of changes in minimum wage laws on levels of employment and \citepos{leighley2013votes} study of whether voter registration laws increase voter turnout. All of these studies share a common structure which favors the use of DID. First, a change in policy is the treatment of interest. Data are then collected in the pre- and post-treatment time periods for a treated group--the place with the new policy--and control group--the place without a change in policy.

\subsection{Motivating sensitivity analysis for differences-in-differences}

Many investigators view the DID method as a credible way to estimate causal effects for two reasons. First, the DID method protects against bias from time-invariant unobserved confounders. Second, the DID device is closely identified with natural experiments, which many authors emphasize as a credible way to estimate treatment effects \citep{Imbens:2010,Rubin:2008,Keele:2015c,rosenbaum2015see,Angrist:2010}. In a natural experiment, the goal is to identify instances where a treatment is assigned through some natural, haphazard process that may approximate as-if random assignment. The DID device is often characterized as a type of natural experiment \citep{mayne2015impact} most likely due to the fact that the DID device has been applied to several well-known examples of natural experiments \citep{card1990impact,freedman1991statistical}. 

A sensitivity analysis asks how strong the effects of an unmeasured covariate would have to be to alter the substantive conclusions from the study. Sensitivity analyses are typically employed in the context of observational studies where treatments are not randomized. A sensitivity analysis for the DID device might seem superfluous given the perceived credibility of this method. However, despite the close association between the DID device and natural experiments, there is nothing about the use of DID that implies a natural experiment. That is, the DID device itself tells you nothing about how treatments were assigned, and in many cases, DID is applied to contexts where treatment assignment is entirely purposeful and thus does not correspond in any way to a natural experiment. For example, in the EDR application, such changes in state policy are not natural experiments in that changes made to voting regulations are purposeful actions by state governments. It is no accident that two upper Midwestern states--Minnesota and Wisconsin--with high levels of voter turnout enacted EDR years before any other state. In fact, the state legislatures in these two states adopted these laws not by accident but to maximize turnout in those states \citep{Smolka:1977}. While analyses that use DID may sometimes give credible evidence, in those cases, it is other aspects of the research design (as-if random treatment assignment, careful data collection, measurement, checks of assumptions) that makes the study credible, not the use of DID on its own. As we highlight below, bias from time-varying confounders is often quite plausible in many analyses that use the DID device. As such, investigators should study whether conclusions based on the DID device are sensitive to possible bias from hidden confounders.

\begin{sidewaysfigure}
  \centering
  \subfloat[Case A]{\includegraphics[width=0.25\textheight]{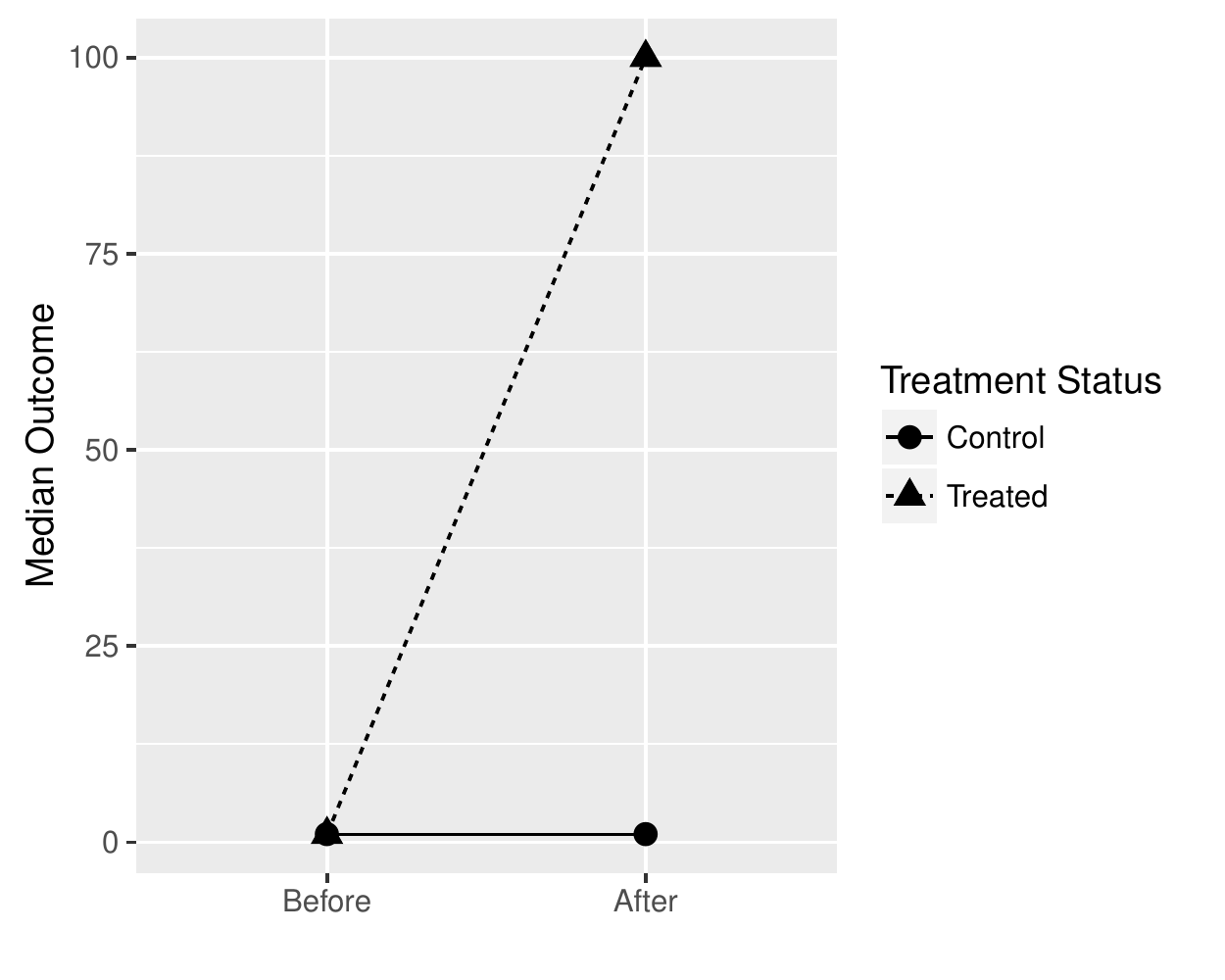}\label{fig:a}}
  \subfloat[Case B]{\includegraphics[width=0.25\textheight]{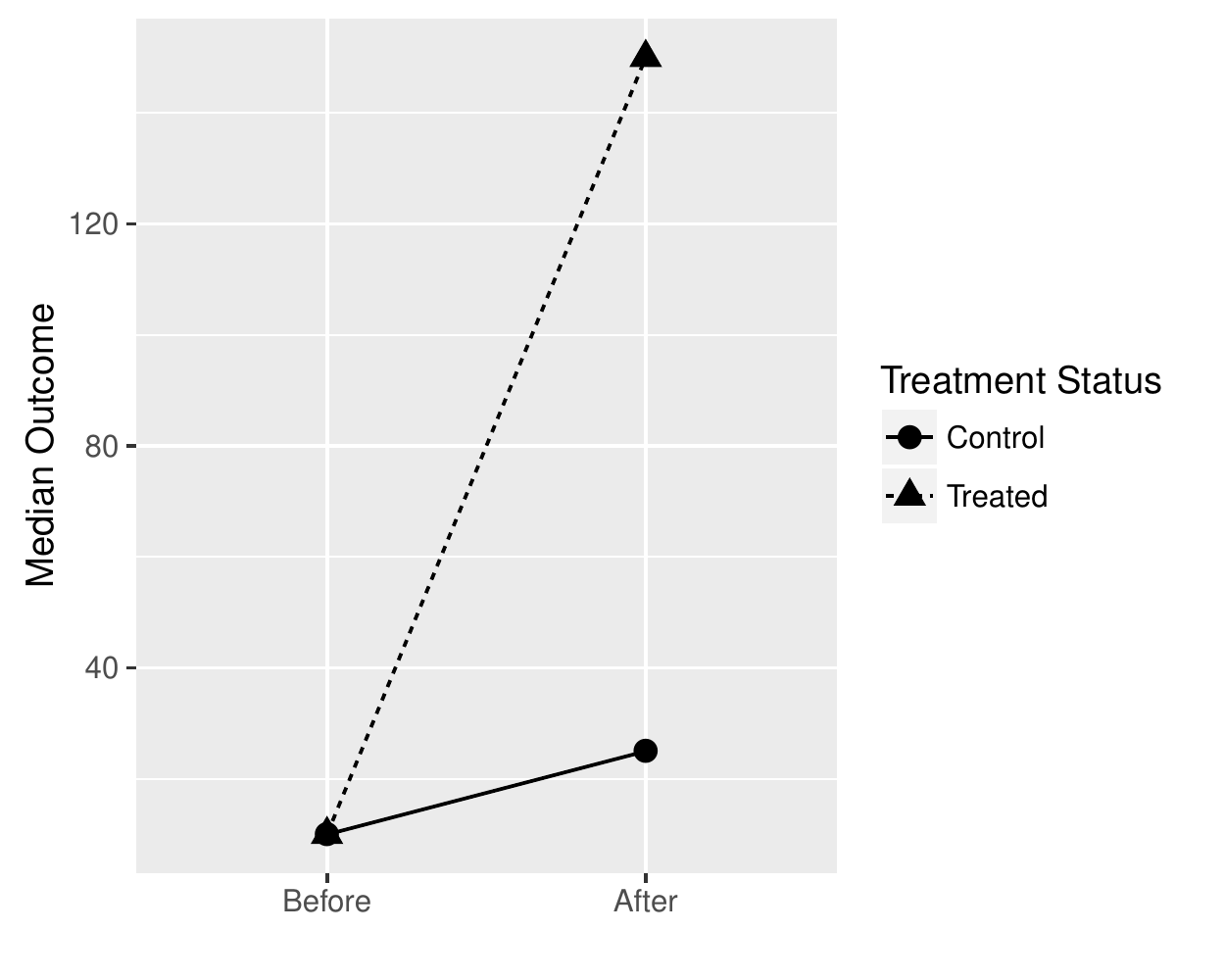}\label{fig:b}}
  \subfloat[Case C]{\includegraphics[width=0.25\textheight]{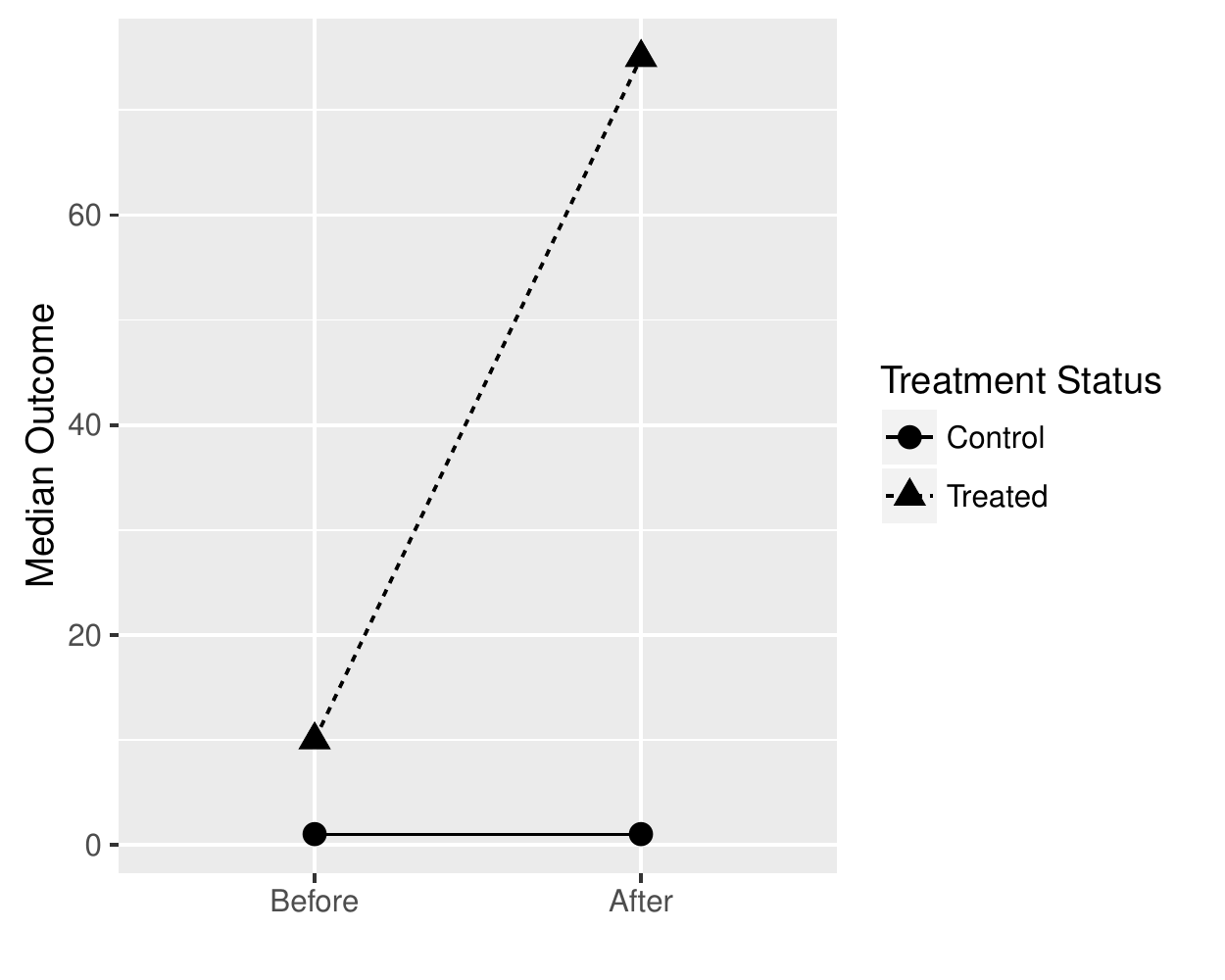}\label{fig:c}}
  \subfloat[Case D]{\includegraphics[width=0.25\textheight]{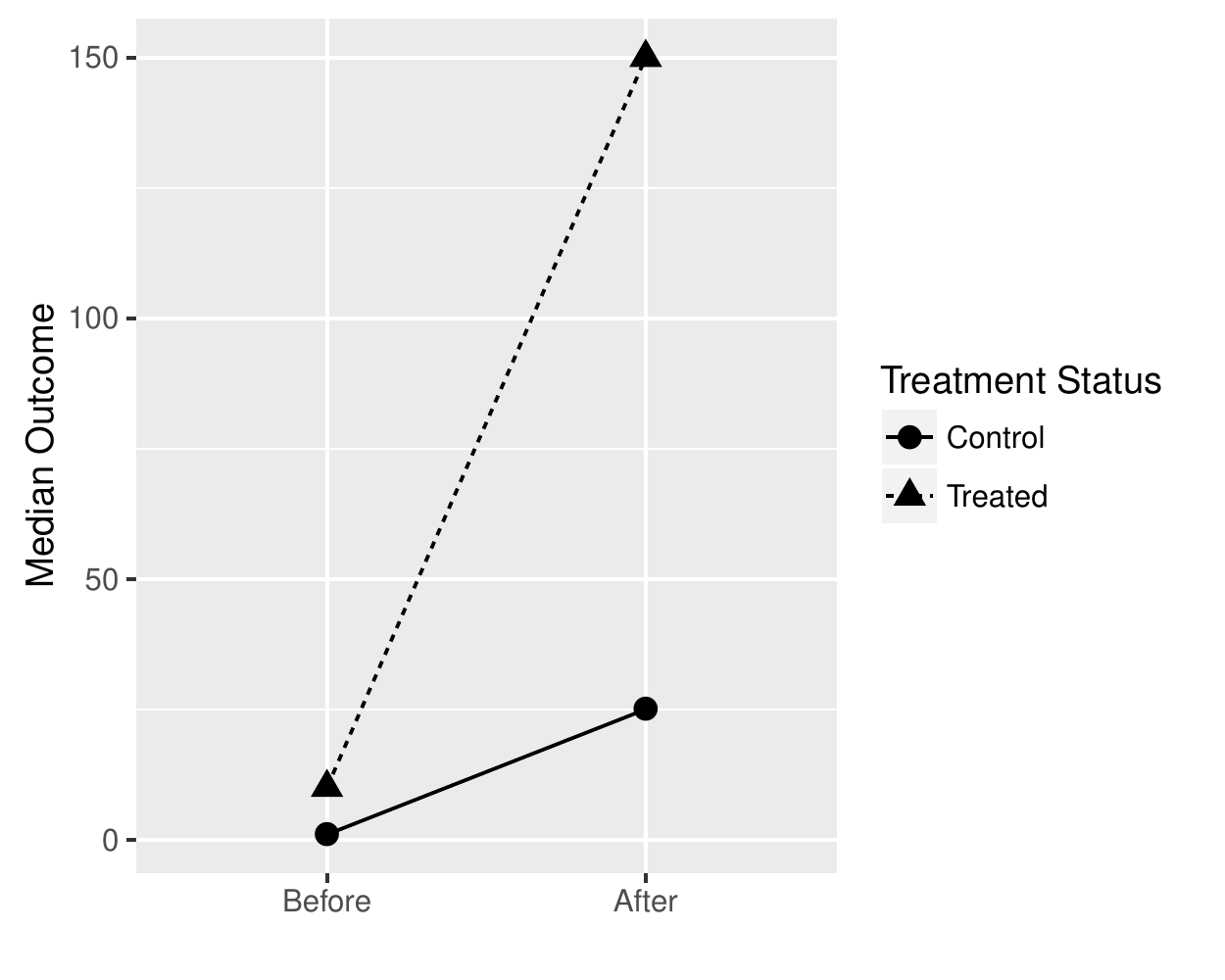}\label{fig:d}}\\
  \subfloat[Case A, Log-Scale]{\includegraphics[width=0.25\textheight]{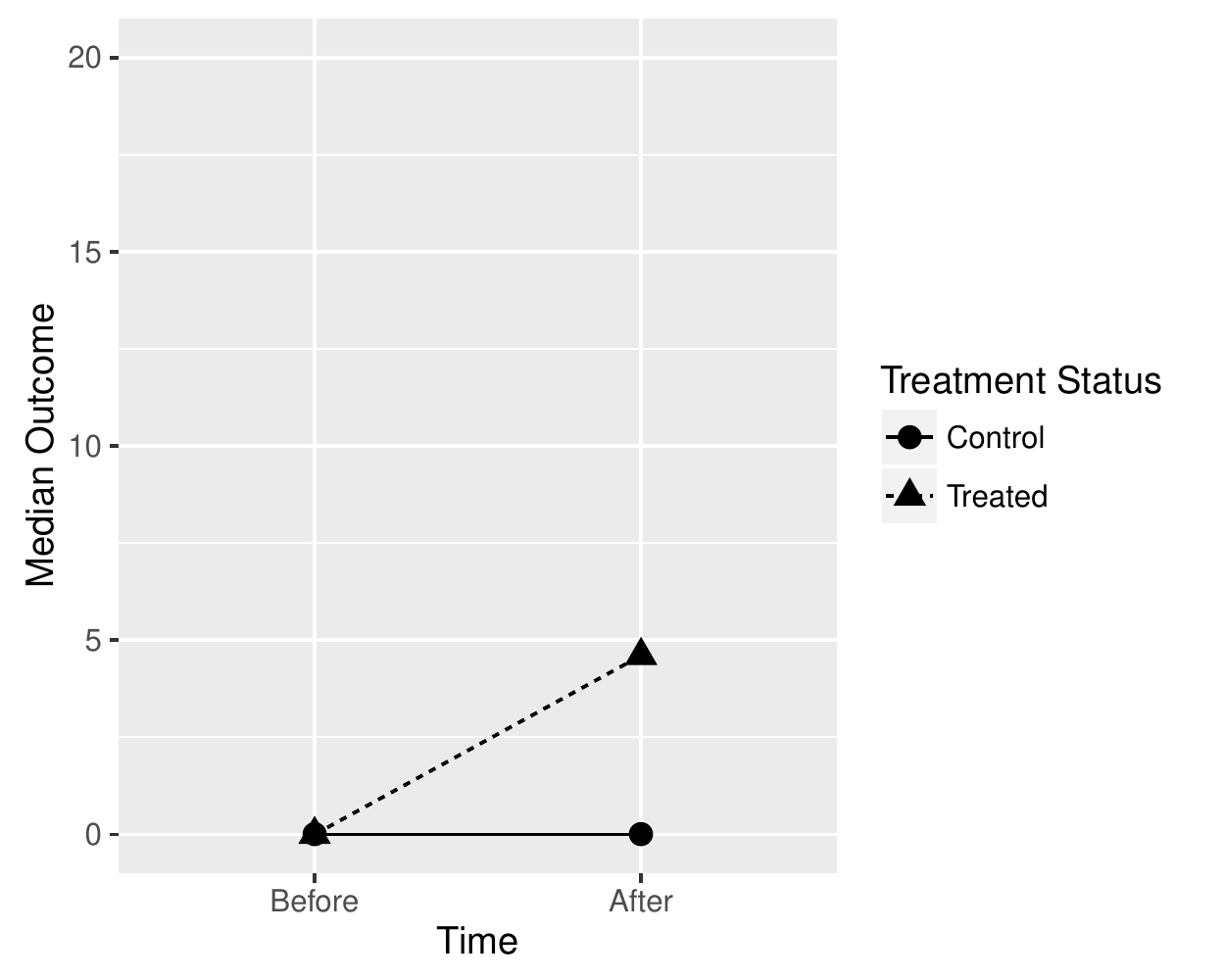}\label{fig:alog}} 
  \subfloat[Case B, Log-Scale]{\includegraphics[width=0.25\textheight]{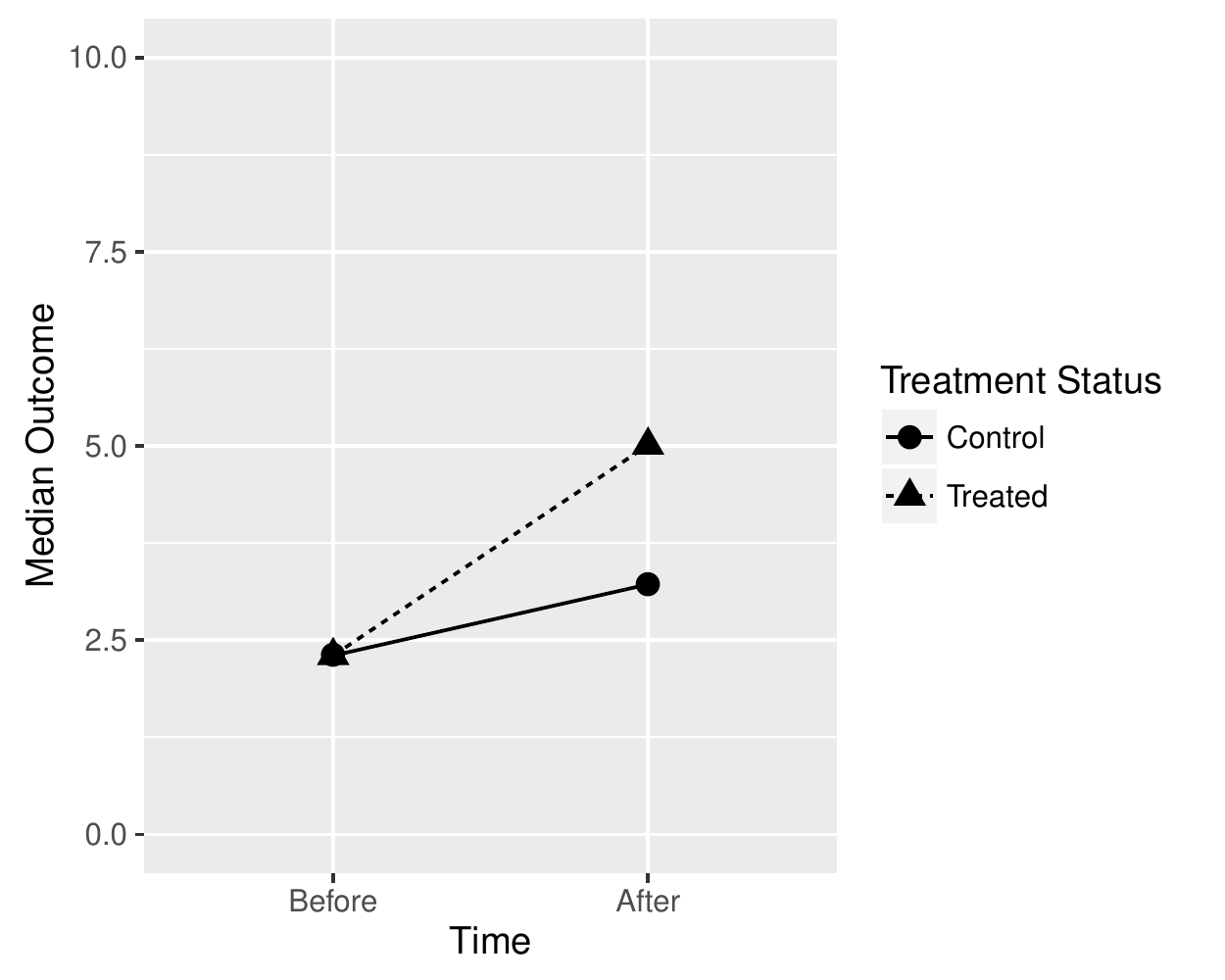}\label{fig:blog}}
  \subfloat[Case C, Log-Scale]{\includegraphics[width=0.25\textheight]{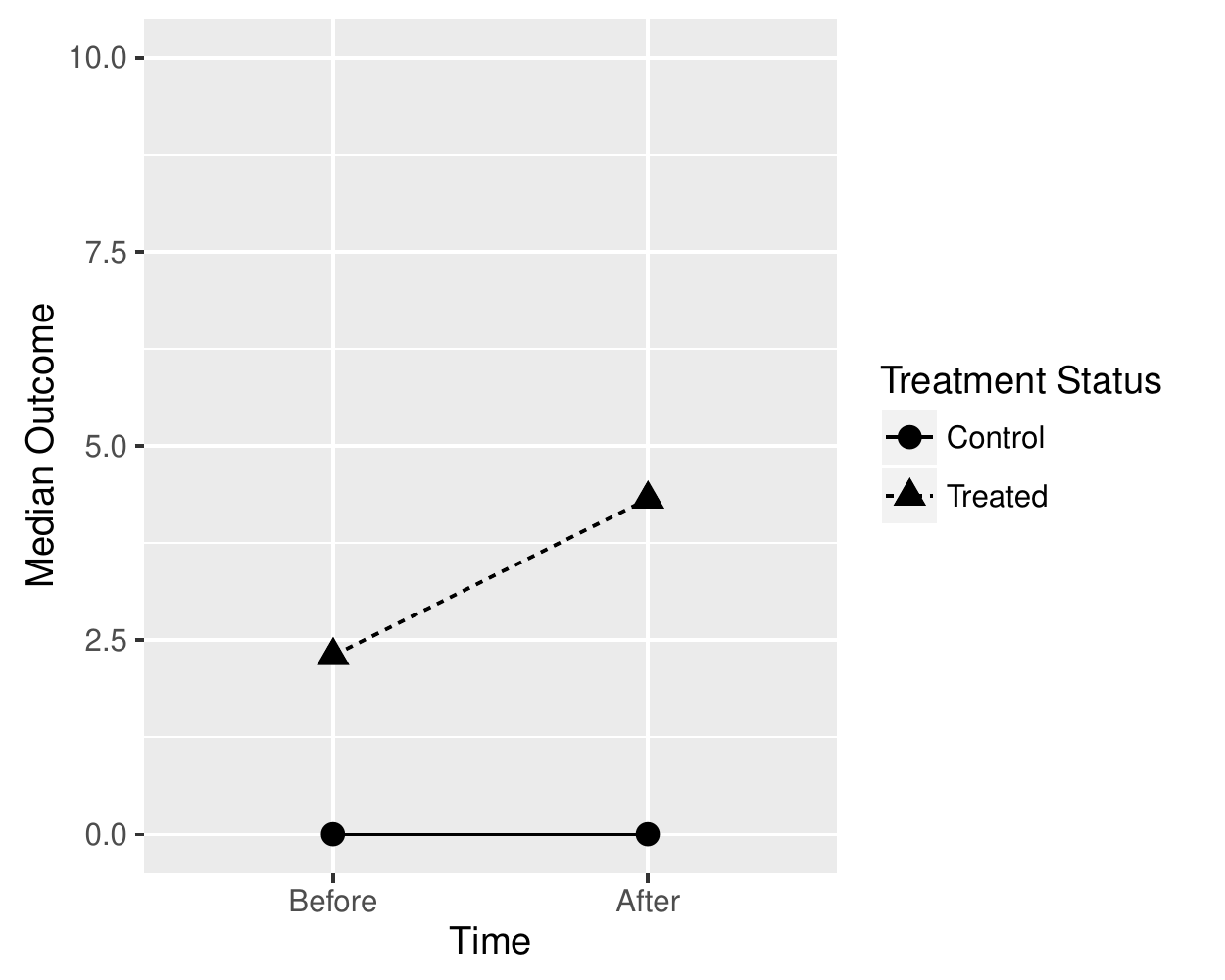}\label{fig:clog}}
  \subfloat[Case D, Log-Scale]{\includegraphics[width=0.25\textheight]{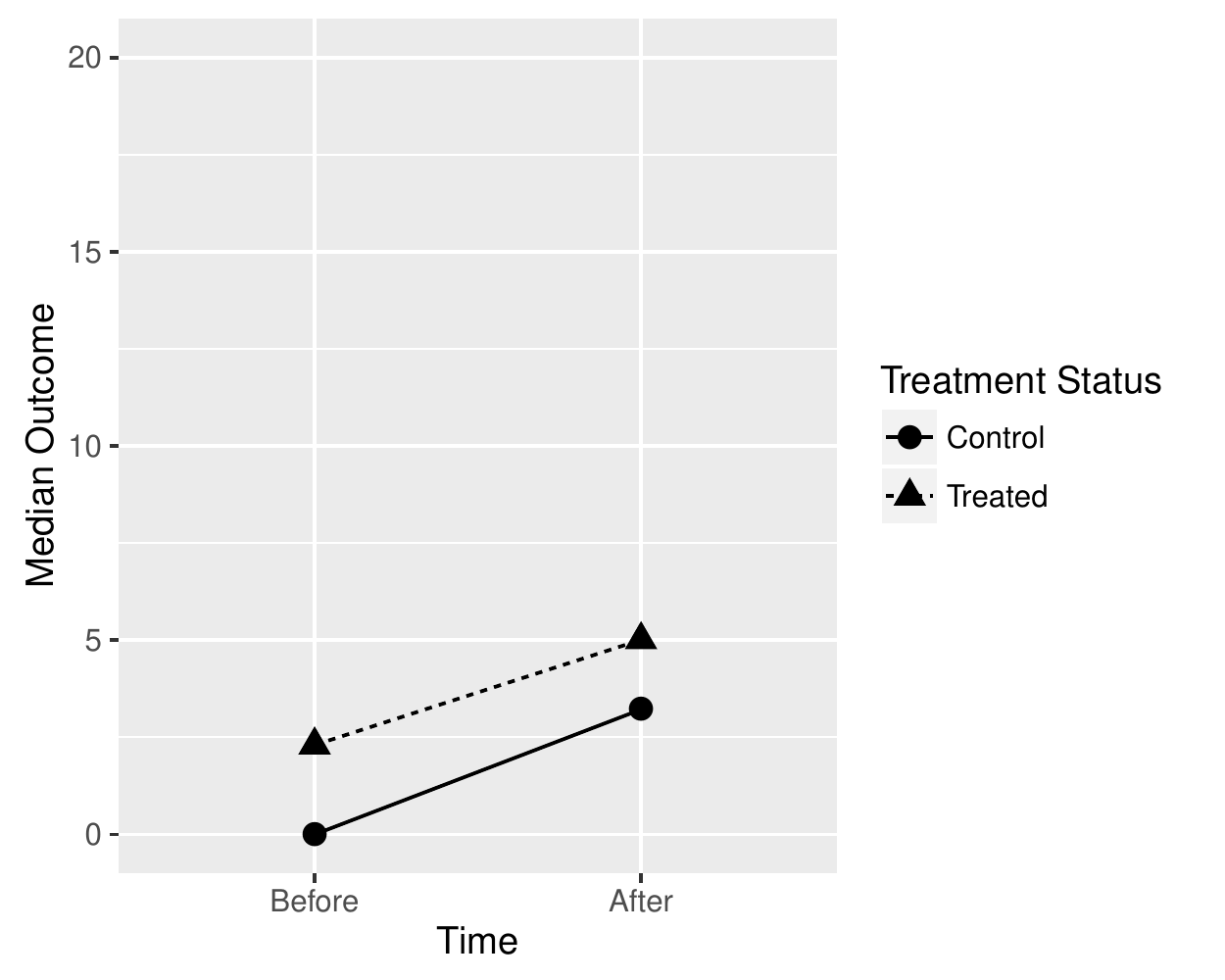}\label{fig:dlog}}
  
\caption{Schematic representation of response in treated and control groups, before and after treatment, with and without transformation to log scale. \label{fig:did}}
\end{sidewaysfigure}

To further explicate this point, we created a graphical demonstration of a discussion from \citet{campbell1969reforms} where he discussed studies of the effects of institutional reforms. Following his discussion, Figure~\ref{fig:did} depicts the median outcome in treated and control groups, in the periods before and after treatment in the treated group. Among the examples in Figure~\ref{fig:did}, case A is the most convincing: treated and control groups had similar outcomes prior to treatment, the control group did not change, but the outcomes increased in the treated group. In case A in Figure~\ref{fig:did}, three different quantities all suggest the same effect of the treatment at the median: the post-treatment difference between treated and control groups, the change from base-line in the treated group, and the interaction or difference-in-differences. Case B is less convincing but not totally unconvincing: treated and control groups had similar outcomes prior to treatment and very different outcomes after treatment, but the control group changed in the absence of treatment, and of course the log transformation changes the magnitudes but not the pattern. In case B, the change from baseline in the treated group is not a plausible estimate because the controls also changed, but the post-treatment difference and the interaction produce the same estimate of effect. Case C is also less convincing than case A, and arguably less convincing than case B: the groups were not comparable prior to treatment, but the treated group changed while the control group did not, and the log transformation changes magnitudes but not the pattern. In case C, the post-treatment difference is not a plausible estimate of effect, but the change in the treated group and the interaction produce the same estimate of effect. Case D is the least convincing, perhaps totally unconvincing: the groups were not comparable prior to treatment, both groups changed, but the treated group changed by a larger amount. Even in the most convincing case, case A, an additional pretreatment measure one period before the plotted pretreatment measure might reveal a lazy X pattern with the cross at the shared before point, so that both groups were on a linear trajectory that did not change after treatment, suggesting no treatment effect.  

In each case, application of differences-in-differences is possible, but absent more detailed background knowledge of how treatments were assigned, there little reason to think of it as a panacea, since the protection against hidden bias offered by differences-in-differences is mostly the result of arithmetic convenience rather than the plausibility that the sole source of bias stems from the additive distortions model. Moreover, these patterns are possible in any instance with longitudinal data, and nothing about these patterns reveal whether treatment assignment might approximate a natural experiment.  

Moreover, the bias reducing properties of the DID device are functional form dependent. Simple transformations of the outcome alter the plausibility of a treatment effect when based on the DID device. The lower portion of Figure~\ref{fig:did} depicts the corresponding situations after a log transformation of the outcome. The log transformation is intended to be just one representative of the family of strictly increasing transformations. Under the log transformation, we observe that the treatment effect is mostly eliminated in case D in Figure~\ref{fig:did}. Therefore, the additive pattern of distorting effects comes and goes with strictly monotone transformations of the response, leading us to doubt that additivity of bias can be the central issue in answering a question about treatment effects. While the DID device offers protection against a specific form of bias, we want to emphasize that observational studies that rely on DID are subject to other forms of bias. Moreover, the pattern that allows investigators to use the DID device does not imply the study is a natural experiment. In general, studies that utilize DID require a sensitivity analysis that allows investigators to characterize whether the study conclusions might easily be explained by hidden bias. 

In this manuscript, we outline a method of sensitivity analysis for applications of differences-in-differences. We first describe the method of differences-in-differences by outlining key notation. We next outline an additive model for bias which is typically assumed to hold when DID is applied. Next, we articulate a covariate adjustment strategy based on matching that can reveal important differences between the treated and control group that may be missed if regression models are used. We then develop a sensitivity analysis for DID.  We develop results for both continuous and binary outcomes. We also demonstrate how a sensitivity analysis for DID can be implemented by modification of existing methods. We also review more general considerations for evaluating DID in the context of natural experiments. Finally, we apply our methods two different empirical applications. In the first, we re-analyze the study of disability payment reforms in Germany \citep{puhani2010}. In the second analysis, we investigate whether EDR increases turnout, using data that are similar to those used in \citet{Hanmer:2009}. In each application, we draw important lessons about how to judge whether an analysis based on DID is likely subject to bias from hidden confounders.

\section{The Method of Differences-in-Differences}

\subsection{Notation} 

We now develop formal notation for studies that use DID. Since DID can be applied to a number of data configurations, we tailor our notation in two ways. First, DID may be used when data is available for more than two time periods. In this case, DID is typically implemented as a linear regression model with two-way fixed effects \citep{Bertrand:2004}. The notation we outline below is tailored to the situation where there are data for just two time points: before and after treatment. If data are available for more than two time points, the data can be collapsed into the two time point data configuration. Alternatively, when data are available for many time points, it can also be analyzed using a design that conditions on the distinctive histories of each unit \citep{Abadie:2010,li2001balanced,zubizarreta2014isolation}. See \citet{oneill2016estimating} for a comparison of DID to conditioning on unit histories. Second, the observed data may consist of the same units over time or may be different units over time. This second configuration of the data often occurs when distinct samples of the treated and control populations are surveyed before and after the treatment goes into effect. Our notation and the methods we develop assume the data are from the second configuration. See \citet{rosenbaum2001stability} for a sensitivity analysis tailored to the case where the same units are observed over time. 

We denote time periods with $t\in \{1,2\}$, where 1 indicates a time period before the treatment and 2 indicates a time period after the treatment has been administered. We assume the formation of $I$ matched quadruples, each with a treated and control individual from time periods 1 and 2 that are matched on $\mathbf{x}_{ij}^{(t)}$ a vector of covariates in quadruple $i$ for the $jth$ individual in the pair from period $t$. After matching, $\mathbf{x}_{i1}^{(1)} = \mathbf{x}_{i2}^{(1)} = \mathbf{x}_{i1}^{(2)} = \mathbf{x}_{i2}^{(2)}$, and distinct quadruplets are independent. For  the moment, we do not comment on how to implement matching with DID; we describe the matching process in more detail in the next section. We let $Z_{ij}^{(t)}$ denote the treatment assignment for individual $ij$ in the sample from period $t\in \{1,2\}$.

Each subject $ij$ in period $t$ has a potential outcome under treatment and control. For instance, individual $ij$ in the period 2 sample would exhibit responses $r_{Tij}^{(2)}$ and $r_{Cij}^{(2)}$ under treatment and control respectively. Because each subject is seen under only one treatment, treatment effects such as $r_{Tij}^{(2)} - r_{Cij}^{(2)}$ are not observed for any subject $ij$; see \citet{Neyman:1923a} and \citet{Rubin:1974}. 

The response actually observed from $ij$ in each time period is $R_{ij}^{(t)}$, and $ 0 \leq u^{(t)}_{ij} \leq 1$ is an unmeasured binary confounder for each individual. Write $\cC = \{x^{(t)}_{ij}, u^{(t)}_{ij}: i=1,...,I; j=1,2; t=1,2\}$, as in \citet{Rosenbaum:2009c}. Write $\Omega = \{\bz: z^{(t)}_{i1}+z^{(t)}_{i2}=1: i=1,..,I; t=1,2\}$, which is the standard restriction for a paired design, and $\cZ = \{\bZ\in \Omega\}$. 

For each set $i$, the interaction or difference-in-differences contrast is
\begin{align}
\label{eq.did}
D_i = (Z_{i1}^{(2)} - Z_{i2}^{(2)})(R_{i1}^{(2)}-R_{i2}^{(2)}) - (Z_{i1}^{(1)} - Z_{i2}^{(1)})(R_{i1}^{(1)}-R_{i2}^{(1)}).
\end{align}
\noindent In the derivations that follow, we focus on testing sharp null hypotheses, i.e. hypotheses which impute the missing values of the potential outcomes. For instance,  Fisher's \citeyearpar{Fisher:1935} sharp null hypothesis of no effect of any kind asserts $r_{Tij}^{(t)} = r_{Cij}^{(t)}$ for all $i,j,t$. While sharp nulls are themselves of scientific interest, a test of the composite null hypothesis that the average of the treatment effects equals some value $\tau_0$ while allowing for heterogeneous effects may also be desired. In \S~\ref{ssec.het}, we demonstrate how investigators can consider inference for average treatment effects through a minor modification of the procedure assuming constant effects.

\subsection{Adjustment for Overt Bias Via Matching}
\label{ssec.match}
When investigators apply the DID device, it is typical to adjust for observed covariates that may be confounders using linear regression models, though see \citet{Abadie:2005,Athey:2006,stuart2014using} for exceptions. As we noted above, we assume a matched structure such that $\mathbf{x}_{i1}^{(1)} = \mathbf{x}_{i2}^{(1)} = \mathbf{x}_{i1}^{(2)} = \mathbf{x}_{i2}^{(2)}$. A match of this form requires that units are balanced both with respect to treatment and control arms, but also with respect to time period. We implement this match by conducting three separate matches. First, we match treated units to control units in the pretreatment time period. This removes possible differences across treated and control groups prior to treatment. Next, we match treated to control units in the post-treatment time period. After these first two matches, we now have two sets of matched pairs, one from the pretreatment time period and one set from the post-treatment time periods. Using these two sets of matched pairs, we next match pretreatment pairs to post-treatment pairs. We will discuss the justification for and benefits of this third match in the next section.

The form of matching in each case need not be specific. Ideally, the matching would be done using an optimization algorithm \citep{Rosenbaum:1989,Ming:2000,Hansen:2004,Zubizarreta:2012}. We implement the matching in the applications below using a method based on integer programming in the \texttt{R} package \texttt{designmatch} \citep{Zubizarreta:2012,zubizarreta2016,kilcioglumaximizing}. This form of matching allows us to specify specific balance constraints for each covariate. As a practical matter, the first two matches are straightforward to implement. The third match, however, requires matching matched pairs to matched pairs, which is not a standard matching problem. For this match, we create a new data set based on summary statistics applied to each set of matched pairs.  That is, we create a dataset from the pretreatment matched pairs and post-treatment matched pairs based on summary statistics. For example, we could use the within pair means to form covariates. We can then apply standard matching methods to match pairs to pairs. In general, the mean may not be a suitable summary within the pairs, especially for nominal covariates. Of course, summary statistics other than the mean may be used to summarize matched pairs. For nominal covariates, we use either exact matching or fine balance to avoid the use of summary statistics. Fine balance constrains an optimal match to exactly balance the marginal distributions of a nominal (or categorical) variable, perhaps one with many levels, placing no restrictions on who is matched to whom. This ensures that no category receives more controls than treated, and so the marginal distributions of the nominal variable are identical between the treatment and control groups. See \citet{Rosenbaum:2007b} and \citet{Yang:2012} for more details on fine balance. If we apply either fine balancing or exact matching to any nominal covariates in the initial matches, we can then exactly match or fine balance these covariates when we match pairs across the two time periods. The end result of this triple matching process are matched sets such that $\mathbf{x}_{i1}^{(1)} \approx \mathbf{x}_{i2}^{(1)} \approx \mathbf{x}_{i1}^{(2)} \approx \mathbf{x}_{i2}^{(2)}$ for all $i$.

One feature of this matching plan is that analysts must carefully select the covariates used in the match. Typically, analysts are advised to only match on covariates measured before the treatment occurs to avoid bias from conditioning on a post-treatment covariate \citep{Rosenbaum:1984b}. For example, in a matched cohort study with data collected on individuals pre- and post-treatment, one would be advised to only match on pretreatment measurements. Our matched plan matches units across time. Thus analysts should match on covariates that are thought to be unaffected by the intervention but may affect the outcome. For example, it is well known that voters with higher levels of education vote at higher levels \citep{Wolfinger:1980}. Thus, under our design, we would want to match on education across both time periods. Education would be a safe covariate for matching, since it is hard to imagine that a change in voter registration laws would affect levels of education.

We advocate matching as a method of adjustment since it allows us to remove overt biases without reference to outcomes. This prevents explorations of the data that may invalidate inferential methods \citep{Rubin:2007}. \citet{imbens2015matching} argues that matching is an attractive alternative to regression models, since it tends to be more robust to a variety of data configurations. The first use of matching in conjunction with the DID device appears to be \citet{Heckman:1998b}.

\subsection{A Model of Additive Bias}
\label{ssec.add}

The DID device is considered useful because it removes two distorting effects when the bias from unobservables follow a specific additive model. These two distorting effects are a uniform time trend, affecting both groups in the same way, which we denote as $\alpha_i$, and a constant group difference between treated and control groups, which we denote $\beta_i$. In this section, we formally express the DID device as a function of these two biases. Under our notation, $\alpha_i$ and $\beta_i$ can vary across matched sets, reflecting potential dependence between these biases and the covariates $x_{ij}^{(t)}$. The matching plan we proposed above is designed to remove bias that stems from the presence of $\beta_i$ that depends on $x_{ij}^{(t)}$.  That is, the matching of pairs across time periods will remove bias due to group status that is thought to vary as a function of $x_{ij}^{(t)}$.

Under these two distorting effects, the potential outcomes under control are generated as
\begin{align}\label{eq.po} (r^{(2)}_{Cij}\mid \cC, Z^{(2)}_{ij}=1) &= \mu_i + \beta_i + \alpha_i + \epsilon^{(2)}(x^{(2)}_{ij},u^{(2)}_{ij})\\
(r^{(2)}_{Cij}\mid \cC, Z^{(2)}_{ij}=0) &= \mu_i + \alpha_i + \epsilon^{(2)}(x_{ij}^{(2)},u^{(2)}_{ij})\nonumber\\
(r^{(1)}_{Cij}\mid \cC, Z^{(1)}_{ij}=1) &= \mu_i + \beta_i + \epsilon^{(1)}(x_{ij}^{(1)},u^{(1)}_{ij})\nonumber\\
(r^{(1)}_{Cij}\mid \cC, Z^{(1)}_{ij}=0) &= \mu_i  + \epsilon^{(1)}(x^{(1)}_{ij},u^{(1)}_{ij}),\nonumber
\end{align} where $\epsilon^{(t)}(x_{ij}^{(t)}, u_{ij}^{(t)})$ are independent across matched sets and drawn from a distribution dependent upon $x_{ij}^{(t)}$, $u_{ij}^{(t)}$ and the time period $t$. If the only aspect of the treatment condition that affected the response was the introduction of the treatment, then $r_{Tij}^{(1)} = r_{Cij}^{(1)}$ for all $ij$, and we refer to this as the hypothesis of an isolated effect of the treatment. An isolated and additive effect $\tau$ of the treatment is that $r_{Tij}^{(2)} - r_{Cij}^{(2)} = \tau$ for all $ij$.  We refer to this as the additive distortions model. When the additive model of bias holds, the interaction or difference-in-difference in (\ref{eq.did}) removes the bias from both of these distorting effects. This encapsulates the key advantage of DID: it removes unobserved additive bias. Often the model of additive bias is referred to as a parallel or common trends assumption \citep[\S 3.2.1]{lechner2011estimation}. That is, when the additive model holds, absent treatment, the over time trends in the treated and control outcomes would follow parallel paths or common trends. It is only the administration of the treatment that moves the treated trends off this parallel path, as reflected above.

Next, we re-write the additive model of bias in a way that allows us to describe both the extent to which it may hold and the ways in which it may fail. In what follows, we will write $\epsilon^{(t)}_{ij}$ for $\epsilon^{(t)}(x^{(t)}_{ij}, u^{(t)}_{ij})$. The long form is useful for encoding the natural assumption that $\epsilon_{ij}^{(t)} \indep Z_{ij}^{(t)} \mid \cC$. That is to say, after accounting for the additive biases, the residuals are assumed independent of treatment assignment conditional upon the measured and unmeasured covariates. 

First, we re-express (\ref{eq.did}) in terms of $\epsilon_{ij}^{(t)}$ as
\begin{align*}
D_i &= (Z_{i1}^{(2)} - Z_{i2}^{(2)})(R_{i1}^{(2)}-R_{i2}^{(2)}) - (Z_{i1}^{(1)} - Z_{i2}^{(1)})(R_{i1}^{(1)}-R_{i2}^{(1)})\\
&= (Z_{i1}^{(2)} - Z_{i2}^{(2)})(r_{Ci1}^{(2)}-r_{Ci2}^{(2)}) - (Z_{i1}^{(1)} - Z_{i2}^{(1)})(r_{Ci1}^{(1)}-r_{Ci2}^{(1)})\\
&= \left\{\mu_i+\alpha_i+\beta_i - (\mu_i + \alpha_i)\right\}- \left(\mu_i+\beta_i - \mu_i\right)  \\
&+ (Z_{i1}^{(2)} - Z_{i2}^{(2)})(\epsilon_{i1}^{(2)}-\epsilon_{i2}^{(2)}) - (Z_{i1}^{(1)} - Z_{i2}^{(1)})(\epsilon_{i1}^{(1)}-\epsilon_{i2}^{(1)})\\
&= (Z_{i1}^{(2)} - Z_{i2}^{(2)})(\epsilon_{i1}^{(2)}-\epsilon_{i2}^{(2)}) - (Z_{i1}^{(1)} - Z_{i2}^{(1)})(\epsilon_{i1}^{(1)}-\epsilon_{i2}^{(1)})
\end{align*}Were it the case that hidden bias only affects the potential outcomes through the additive form above as is commonly assumed in applications of DID, we may further assume that $\epsilon^{(t)}(x^{(t)}_{ij}, u^{(t)}_{ij}) = \epsilon^{(t)}(x^{(t)}_{ij})$. As $x^{(t)}_{ij}$ is assumed the same for all individuals in the quadruplet after matching, this would imply that $\epsilon_{i1}^{(1)}$ and $\epsilon_{i2}^{(1)}$ would be $iid$ from the time 1 distribution, while $\epsilon_{i1}^{(2)}$ and $\epsilon_{i2}^{(2)}$ would be $iid$ from the time 2 distribution. As a result, $(\epsilon_{i1}^{(t)} - \epsilon_{i2}^{(t)})$ would be symmetrically distributed about zero for $t=1,2$, and would be equal in distribution to $(\epsilon_{i2}^{(t)} - \epsilon_{i1}^{(t)})$ for $t=1,2$. By ignorability, $\E[(Z_{i1}^{(t)} - Z_{i2}^{(t)})(\epsilon_{i1}^{(t)}-\epsilon_{i2}^{(t)})\mid \cC] = 0$ for $t=1,2$, and the distribution of the observed differences-in-differences would also be symmetric about zero. This ideal setting can be expressed through two equations as, for any $w>0$,

\begin{align*}
\P\left\{\epsilon_{i1}^{(2)} - \epsilon_{i2}^{(2)} - (\epsilon_{i1}^{(1)} - \epsilon_{i2}^{(1)}) > w\mid \cC\right\} &= \P\left\{\epsilon_{i1}^{(2)} - \epsilon_{i2}^{(2)} - (\epsilon_{i1}^{(1)} - \epsilon_{i2}^{(1)}) <-w\mid \cC\right\}\\
\P\left\{\epsilon_{i1}^{(2)} - \epsilon_{i2}^{(2)} - (\epsilon_{i2}^{(1)} - \epsilon_{i1}^{(1)}) > w\mid \cC\right\} &= \P\left\{\epsilon_{i1}^{(2)} - \epsilon_{i2}^{(2)} - (\epsilon_{i2}^{(1)} - \epsilon_{i1}^{(1)}) < -w\mid \cC\right\}
\end{align*}  
Note the difference in the two lines here: the swapping of the indices in period 1. Next, we use the additive model of bias to develop a more general sensitivity analysis for the DID device.

\section{Sensitivity Analysis for DID}
\label{sec.sens}

\subsection{Bounds for Hidden Biases of Two Forms}
\label{ssec.twoparam}
We now develop a sensitivity analysis for quantities of interest based on DID. A sensitivity analysis allows an investigator to quantify the degree to which a key assumption must be violated in order for the study conclusions to be reversed. If an inference is sensitive, a slight violation of the assumption may lead to substantively different conclusions. Our model extends the framework of \citet{Rosenbaum:2009c} to the case of differences-in-differences. The model simultaneously bounds the potential impact of hidden bias due to departures from the model of additive bias and the differences in treatment status. We further establish a mapping known as an amplification between the resulting sensitivity analysis and a model which only bounds the odds of assignment to treatment as in \citet[ch. 4]{Rosenbaum:2002}. Through this sensitivity analysis, one attains bounds on quantities such as the treatment effect point estimate or p-value based upon the degree to which hidden bias is allowed to corrupt the inference.

If hidden bias exists, then the distribution of the difference in these residuals need not be symmetric due to the impact of unmeasured confounders $u_{ij}^{(t)}$ on the distribution of $\epsilon_{ij}^{(t)}$. For example, bias could result from departures from the additive group differences and/or additive time trend assumptions that underlie the DID device. A natural way to express the impact of hidden bias is through departures from symmetry in the differences-in-differences of the residuals. We use the following model to describe departures from symmetry in the residuals. The model has two parameters for periods 1 and 2, $\delta_1$ and $\delta_2$ respectively. For any $w>0$,
\begin{align}
\label{eq:delta}
\P\left\{\epsilon_{i1}^{(2)} - \epsilon_{i2}^{(2)} - (\epsilon_{i1}^{(1)} - \epsilon_{i2}^{(1)}) > w\mid \cC\right\} &= \frac{\exp\{\delta_2(u^{(2)}_{i1}-u^{(2)}_{i2})\}}{\exp\{\delta_1(u^{(1)}_{i1}-u^{(1)}_{i2})\}}\P\left\{\epsilon_{i1}^{(2)} - \epsilon_{i2}^{(2)} - (\epsilon_{i1}^{(1)} - \epsilon_{i2}^{(1)}) < -w\mid \cC\right\}\\ 
\P\left\{\epsilon_{i1}^{(2)} - \epsilon_{i2}^{(2)} - (\epsilon_{i2}^{(1)} - \epsilon_{i1}^{(1)}) > w\mid \cC\right\} &= \frac{\exp\{\delta_2(u^{(2)}_{i1}-u^{(2)}_{i2})\}}{\exp\{\delta_1(u^{(1)}_{i2}-u^{(1)}_{i1})\}}\P\left\{\epsilon_{i1}^{(2)} - \epsilon_{i2}^{(2)} - (\epsilon_{i2}^{(1)} - \epsilon_{i1}^{(1)}) < -w\mid \cC\right\} \nonumber
\end{align}

Under this model, $\delta_1=\delta_2=0$ corresponds to the case of symmetry or the case where the additive model holds, and the DID device successfully removes any hidden bias.  Positive or negative values of $\delta_1$ or $\delta_2$ allow for departures from the additive model of bias and for different strengths and directions of correlations between the unmeasured confounders and the odds of observing a positive difference in residuals relative to a negative one. This model allows for biases that ``cancel out'' due to happenstance, but also allows for the amplification of biases due to time period interactions. 

Under the DID device, bias might also arise due to differential probability of treatment assignment. We further characterize the unknown probability that a subject is exposed to treatment using the following model:
\begin{align}
\label{eq.trt}
\P(\bZ=\bz\mid \cZ, \cC) &= \prod_{i=1}^I\prod_{t=1}^{2}\frac{\exp\{\lambda_t(z^{(t)}_{i1}u^{(t)}_{i1}+z^{(t)}_{i2}u^{(t)}_{i2})}{\exp(\lambda_t u^{(t)}_{i1}) + \exp(\lambda_tu^{(t)}_{i2})}
\end{align}
where $\lambda_1$ and $\lambda_2$ are period-specific parameters and $\lambda_1=\lambda_2=0$ encodes the case where a hidden confounder does not alter the probability of assignment. Next, we construct sharp worst-case bounds for quantities of interest for continuous outcomes as a function of possible departures from these two models for hidden bias. Similar, though distinct derivations, for binary outcomes follow in \S \ref{ssec.bin}. 

Critically, the key independence assumption under the DID device is made at the level of the residuals, i.e. that $\epsilon^{(t)}_{ij}\indep Z^{(t)}_{ij}\mid \cC$ instead of at the level of potential outcomes, i.e. $r^{(t)}_{Cij}\indep Z^{(t)}_{ij}\mid \cC$. As such, development of a sensitivity analysis requires employing the distribution for these residuals, rather than of the original potential outcomes under control. This is due to the fact that the observed DID contrast provides a value for $(Z_{i1}^{(2)} - Z_{i2}^{(2)})(\epsilon_{i1}^{(2)}-\epsilon_{i2}^{(2)}) - (Z_{i1}^{(1)} - Z_{i2}^{(1)})(\epsilon_{i1}^{(1)}-\epsilon_{i2}^{(1)})$. For instance, if the first individual received the treatment in each pair, we observe $r^{(2)}_{Ci1}-r^{(2)}_{Ci2} - (r^{(1)}_{Ci1} - r^{(1)}_{Ci2})  = (\epsilon_{i1}^{(2)}-\epsilon_{i2}^{(2)}) - (\epsilon_{i1}^{(1)}-\epsilon_{i2}^{(1)})$. In contrast to the more conventional setting where Fisher's sharp null imputes $r^{(t)}_{Cij}$ for all individuals, we do not know $\epsilon_{ij}^{(t)}$ for any individual even under the sharp null, nor do we observe $\epsilon_{i1}^{(t)}-\epsilon_{i2}^{(t)}$ for any $j$ or $t$. Instead, from the observed we know the value for either $\epsilon_{i1}^{(2)} - \epsilon_{i2}^{(2)} - (\epsilon_{i1}^{(1)} - \epsilon_{i2}^{(1)})$ or $\epsilon_{i1}^{(2)} - \epsilon_{i2}^{(2)} - (\epsilon_{i2}^{(1)} - \epsilon_{i1}^{(1)})$ depending on the treatment assignment under the sharp null, but one does not provide the other. 

This limited information necessitates conducting inference over a further subset of $\Omega$. To proceed, we first condition on $\cZ$ (i.e. the paired design) and on the values $\bB = (B_1,...,B_I)^T$, where $B_i = (Z_{i1}^{(1)}-Z_{i2}^{(1)})(Z_{i1}^{(2)}- Z_{i2}^{(2)})$. For each pair, this restricts attention to (1) the observed treatment assignment and (2) the assignment where, within each quadruple, we simultaneously swap the treatment assignments for the pair in period $1$ and the pair in period $2$. This eliminates consideration of hypothetical assignments where only one treatment is flipped in a quadruple relative to what was actually observed. 

Recall that $\epsilon_{ij}^{(t)} \indep Z_{ij}^{(t)} \mid \cC$ by assumption. As $(\cZ, \bB)$ is a function of $\bZ$ and we have independence across quadruplets, it follows that $\epsilon_{ij}^{(t)}\indep Z^{(t)}_{ij} \mid \cC, \cZ, \bB$ \citep[][Lemma 4.2]{dawid1979conditional}. Defining $V_i = (Z^{(2)}_{i1}-Z^{(2)}_{i2})$,  the observed difference-in-difference contrast can be expressed as 
\begin{align*} D_i &= V_i(\epsilon_{i1}^{(2)}-\epsilon_{i2}^{(2)}) - V_iB_i(\epsilon_{i1}^{(1)}-\epsilon_{i2}^{(1)}). \end{align*}
Fix $B_i=b_i$ by conditioning, and observe that $(\epsilon_{i1}^{(2)}-\epsilon_{i2}^{(2)}) - b_i(\epsilon_{i1}^{(1)}-\epsilon_{i2}^{(1)} )\indep \bZ_i \mid \cC, \cZ, \bB$. Let $Y_i=\left(\epsilon_{i1}^{(2)}-\epsilon_{i2}^{(2)} - b_i(\epsilon_{i1}^{(1)}-\epsilon_{i2}^{(1)})\right)$, $A_i = |Y_i|$, and $S_i = \sign(Y_i)$. 

For a fixed set of sensitivity parameters $\delta_1, \delta_2, \lambda_1, \lambda_2$, we now construct a sensitivity analysis based on worst-case bounds for any test statistic of the form $\sum_{i=1}^IV_iS_iq_i$ for some non-negative function $q_i = q(\mathbf{A})_i$ taking the value zero if $A_i = 0$. Most commonly encountered test statistics such as Wilcoxon's signed rank test or the permutational $t$-test are either of this form or are linear functions of this form. The distributional form given in (\ref{eq:delta}) is a necessary and sufficient condition for the sign of a random variable to be independent of its magnitude \citep[][Theorem 2.1]{wolfe1974characterization}. Let $\eta_{si} = \P(S_i = s \mid \cC, \cZ, \bB)$. From (\ref{eq:delta}), we have for $Y_i\neq 0$ 
\begin{align*}
\eta_{1i} &= \frac{\exp\{\delta_2(u^{(2)}_{i1}-u^{(2)}_{i2}) - b_i\delta_1(u^{(1)}_{i1}-u^{(1)}_{i2})\}}{1+\exp\{\delta_2(u^{(2)}_{i1}-u^{(2)}_{i2}) - b_i\delta_1(u^{(1)}_{i1}-u^{(1)}_{i2})\}}\\
\eta_{-1i} &= 1-\eta_{1i},
\end{align*}
while if $Y_i=0$ then $\eta_{0i} = 1$. Next, we write $\rho_i = \P(V_i=1 \mid \cC, \cZ, \bB$). 
\begin{align*}
\rho_i &= \frac{\exp\{\lambda_2(u^{(2)}_{i1}-u^{(2)}_{i2}) + b_i\lambda_1(u^{(1)}_{i1}-u^{(1)}_{i2})\}}{1+\exp\{\lambda_2(u^{(2)}_{i1}-u^{(2)}_{i2}) + b_i\lambda_1(u^{(1)}_{i1}-u^{(1)}_{i2})\}}.\\
\end{align*} 
As $\bY\indep \bZ\mid \cC, \cZ, \bB$ and $\bA$ is a function of $\bY$, it follows that $\bA \indep \bZ\mid \cC, \cZ, \bB$  \citep[][Lemma 4.2]{dawid1979conditional}. Further by (\ref{eq:delta}) $\bA\indep \bS\mid \cC, \cZ, \bB$ \citep[][Theorem 2.1]{wolfe1974characterization}. Consequently, $\bS \indep \bZ \mid \bA, \cC, \cZ, \bB$ \citep[][Lemma 4.3]{dawid1979conditional}. That is, $\rho_i = \P(V_i=1 \mid \cC, \cZ, \bB) = \P(V_i=1 \mid \bA, \cC, \cZ, \bB, \bS)$. Hence,
\begin{align}
\label{eq:prob}
&\P(S_iV_i = 1 \mid \bA, \cC, \cZ, \bB) \\\nonumber
& = \rho_i\eta_{1i} + (1-\rho_i)\eta_{-1i}\\\nonumber
&= \frac{\exp\{(\lambda_2+\delta_2)(u^{(2)}_{i1}-u^{(2)}_{i2}) + b_i(\lambda_1-\delta_1)(u^{(1)}_{i1}-u^{(1)}_{i2})\}+1}{[1+\exp\{\lambda_2(u^{(2)}_{i1}-u^{(2)}_{i2}) + b_i\lambda_1(u^{(1)}_{i1}-u^{(1)}_{i2})\}][1+\exp\{\delta_2(u^{(2)}_{i1}-u^{(2)}_{i2}) - b_i\delta_1(u^{(1)}_{i1}-u^{(1)}_{i2})\}]}.
\end{align}
Suppose $1/\Lambda \leq \exp\{\lambda_1\}, \exp\{\lambda_2\} \leq \Lambda$, and that $1/\Delta \leq \exp\{\delta_1\}, \exp\{\delta_2\} \leq \Delta$. It is straightforward to show that given $u_{ij}^{(t)}\in [0,1]$ the following sharp bounds hold:
\begin{align}\label{eq:bounds}
\frac{\Delta^2+\Lambda^2}{(1+\Lambda^2)(1+\Delta^2)}\leq \P(S_iV_i = 1 \mid \bA, \cC, \cZ, \bB) \leq \frac{(\Lambda\Delta)^2+1}{(1+\Lambda^2)(1+\Delta^2)}.
\end{align}
Attaining these worst-case bounds requires a misalignment of the signs of either $\lambda_1$ and $\lambda_2$ or $\delta_1$ and $\delta_2$. This is apparent in the numerator of (\ref{eq:prob}): the first difference in unmeasured confounders is multiplied by $\lambda_2+\delta_2$, while the second is multiplied by $\lambda_1-\delta_1$. This reflects the fact that when the DID device is employed, it might be susceptible to bias from a hidden confounder and how the treatment and/or outcome evolve over time.

These bounds also allow the investigator to specify specific patterns in the hidden confounding and the evolution of the treatment or outcome between the two time periods. For instance, one could stipulate that $u_{i1}^{(1)} - u_{i2}^{(1)} > 0$ increases both the probability of assignment and the probability of a positive difference in period 1, and this holds in period 2 as well. In terms of the parameters, this is equivalent to $\text{sign}(\delta_1\delta_2) = 1$, and $\text{sign}(\lambda_1\lambda_2) = 1$. Under this assumption, one gets tighter bounds on $\P(S_iV_i = 1 \mid A, \cC, \cZ, \bB)$ for a given value of $\Lambda$ and $\Delta$.  

\subsection{Tying it Together: A New Amplification}
\label{ssec.amp}

The model described in \S \ref{ssec.twoparam} presented probability bounds in terms of two parameters, $\Lambda$ and $\Delta$. Alternatively, the probability bounds may also be written in terms of a single parameter. As we now describe, the two-parameter model based upon $(\Lambda, \Delta)$ provides an \textit{amplification} of a one-parameter model. An amplification is a mapping of a one-parameter sensitivity analysis to a curve of two-parameter sensitivity analyses. Comparing the bounds (\ref{eq:bounds}) to (\ref{eq:gsq}) provides the relationship
\begin{align*}
\Gamma^2 = \frac{\Delta^2\Lambda^2+1}{(\Lambda^2 + \Delta^2)}.
\end{align*}
 \noindent Here, no constraints are imposed on how the direction of the relationship between unmeasured confounders and both the treatment assignment and the potential outcomes changes over time. Next, we write the bounds in terms of a single parameter that we denote as $\Gamma$. For a given $\Gamma$, the amplification set in the DID design is $\mathcal{A}_{\Gamma,DID} = \{(\Lambda,\Delta): \Gamma^2 = (\Lambda^2\Delta^2+1)/(\Lambda^2+\Delta^2)\}$. T
 
Compare the amplification for DID to the usual amplification for a paired design developed in \citet{Rosenbaum:2009c}, which is based on the relationship
\begin{align*}
\Gamma = \frac{\Delta\Lambda+1}{\Lambda+\Delta}.
\end{align*}
This implies that despite that despite the equivalence between a sensitivity analysis for a paired design at $\Gamma^2$ and the DID device at $\Gamma$, the maps from the two parameter case to the one parameter case are not identical. For instance, at $\Gamma=2$ then $\mathcal{A}_{2, Pair}$ includes $(\Lambda, \Delta) = (3,5)$.  Contrast that with $\mathcal{A}_{2, DID}$, which instead has the pair $(\Lambda, \Delta) = (3,\sqrt{35/5})\approx (3, 2.65)$. 

For the DID design, a different amplification set would be attained if one required $\delta_1=\delta_2$ and $\lambda_1=\lambda_2$. For fixed $\Gamma$ and $\Lambda$, the corresponding $\Delta$ would be larger. For fixed $\Gamma$ and $\Delta$, the corresponding $\Lambda$ would be larger. As we demonstrate in the next section, expressing the sensitivity analysis in terms of $\Gamma$ allows investigators to use conventional methods to conduct a DID sensitivity analysis. Alternatively, a researcher can proceed with the enriched interpretations from the two-parameter model in \S \ref{ssec.twoparam}.

\subsection{A Conventional Sensitivity Analysis Bounding Only Treatment Assignment}
\label{ssec.conventional}

Next, we present an alternative, though closely related, method for a sensitivity analysis for DID that can be implemented using conventional methods for paired studies as described in \citet[ch. 4]{Rosenbaum:2002}. Let $\mathcal{F}_\epsilon = \{x^{(t)}_{ij}, u_{ij}^{(t)}, \epsilon_{ij}^{(t)}: i=1,...,I; j=1,2; t=1,2\}$ be the set containing the measured and unmeasured covariates but containing the residuals $\epsilon_{ij}^{(t)}$ in place of the potential outcomes found in the conventional model of \citet[ch. 4]{Rosenbaum:2002}. Conditional on $\mathcal{F}_\epsilon$, we write the following model for treatment assignment:
\begin{align}\label{eq:sensconv}
\log\left(\frac{\P(Z_{ij}^{(t)} = 1 \mid \cF_\epsilon)}{\P(Z_{ij}^{(t)}=0\mid \cF_\epsilon)}\right) &= \kappa_t(x^{(t)}_{ij}) + \gamma_tu^{(t)}_{ij},
\end{align}
As before, Fisher's sharp null does not impute the values of $\epsilon_{ij}^{(t)}$, such that the contents of $\mathcal{F}_\epsilon$ are not entirely specified under the null. This again necessitates conditioning on a subset of treatment assignments wherein inference can proceed. Defining $B_i$ and $V_i$, as before, we proceed with inference conditional upon $\cZ$ and $\bB$. Consider now the conditional probability for a given vector of treatment assignments $\bZ_i = (Z_{i1}^{(1)}, Z_{i2}^{(1)}, Z_{i1}^{(2)}, Z_{i2}^{(2)})$. Under these conditions $\bZ_i$ can only take two values. If $b_i=1$, then the values are $(1, 0,1,0)$ or $(0, 1,0,1)$. If $b_i=-1$, the values are $(1, 0,0,1)$ or $(0,1,1,0)$. In general, conditional probability of treatment assignment has the form
\begin{align*}
\P(\bZ_i = \bz_i\mid  \cZ, \cF_\epsilon, \bB=\mathbf{b})
&= \frac{\exp\{\gamma_1(z_{i1}^{(1)} - z_{i2}^{(1)})(u_{i1}^{(1)}-u_{i2}^{(1)}) + \gamma_2b_i(z_{i1}^{(2)}-z_{i2}^{(2)})(u_{i1}^{(2)}-u_{i2}^{(2)})\}}{1+ \exp\{\gamma_1(z_{i1}^{(1)} - z_{i2}^{(1)})(u_{i1}^{(1)}-u_{i2}^{(1)} )+ \gamma_2b_i(z_{i1}^{(2)}-z_{i2}^{(2)})(u_{i1}^{(2)}-u_{i2}^{(2)})\}}.
\end{align*}

If $1/\Gamma \leq \exp(\gamma_1), \exp(\gamma_2) \leq \Gamma$,
\begin{align}
\label{eq:gsq} 
\frac{1}{1+\Gamma^2}\leq\P(\bZ_i = \bz_i\mid \cZ, \cF_\epsilon, \bB)\leq \frac{\Gamma^2}{1+\Gamma^2},
\end{align}
which provides the bound $1/(1+\Gamma^2)\leq \P(V_i=1\mid \cZ, \cF_\epsilon, \bB) \leq \Gamma^2/(1+\Gamma^2)$. Let $y_i = \epsilon_{i1}^{(2)}-\epsilon_{i2}^{(2)} - b_i(\epsilon_{i1}^{(1)}-\epsilon_{i2}^{(1)})$. Under Fisher's sharp null and conditional upon $\cF_\epsilon$, $y_i$ is fixed across the subset of $\Omega$ such that $\mathbf{B} = \mathbf{b}$. Given $\cF_\epsilon$, $\cZ$, and $\mathbf{B}$, the randomization distribution for any test statistic of the form $T = \sum_{i=1}^I q(|\mathbf{y}|)\text{sign}(V_iy_i)$ can then be employed. 

This implies that investigators can conduct a sensitivity analysis under model (\ref{eq:sensconv}) using classical sensitivity analysis methods at $\Gamma^2$ if $\exp(|\gamma_1|), \exp(|\gamma_2|)\leq \Gamma$. Here, we briefly elaborate this point. The usual application of a sensitivity analysis based on Rosenbaum's method consists of applying a statistic such as Wilcoxon's signed rank to treated and control matched pairs. For a given value of $\Gamma$, upper and lower bounds for quantities such a p-values and confidence intervals may be derived for this test statistic. Our result demonstrates that for the DID device, the investigator can apply a statistic such Wilcoxon's signed rank to the DID contrast, and the bounds for $\Gamma$ can be calculated using existing methods but with $\Gamma^{2}$ replacing $\Gamma$. Alternatively, investigators often seek to identify the value or changepoint in $\Gamma$ where a quantity like a p-value is no longer significant. As such, when an analysis is based on DID, one can use conventional methods for paired studies at $\Gamma^{2}$ to find the changepoint value of $\Gamma$.

\subsection{Differences-in-Differences with Heterogeneous Effects}
\label{ssec.het}

Our developments thus far have focused on tests of sharp null hypotheses for treatment effects, such as Fisher's sharp null and a model of constant effects to describe response to treatment. One might be concerned that a sensitivity analysis based on constant effects could be misleading if treatment effects are instead heterogeneous. For paired observational studies, \citet{fogary2017avg} recently developed a sensitivity analysis for the sample average treatment effect while allowing for effect heterogeneity by means of an studentization argument. The resulting sensitivity analysis is exact if the treatment effects are constant at some value $\tau_0$, while asymptotically correct if the effects merely average to $\tau_0$. As we now demonstrate, the methods developed therein extend to the type of matched differences-in-differences analysis that we focus on.

First, we modify the period $2$ response functions given in (\ref{eq.po}) to be
\begin{align*} (r^{(2)}_{Cij}\mid \cC, Z^{(2)}_{ij}=1) &= \mu_i + \beta_i + \alpha_i + \epsilon^{(2)}_{Cij}\\
(r^{(2)}_{Cij}\mid \cC, Z^{(2)}_{ij}=0) &= \mu_i + \alpha_i + \epsilon^{(2)}_{Cij}\\
(r^{(2)}_{Tij}\mid \cC, Z^{(2)}_{ij}=1) &= \mu_i + \beta_i + \alpha_i + \epsilon^{(2)}_{Tij}\\
(r^{(2)}_{Tij}\mid \cC, Z^{(2)}_{ij}=0) &= \mu_i + \alpha_i + \epsilon^{(2)}_{Tij},
\end{align*}and observe that $\tau^{(2)}_{ij} := r^{(2)}_{Tij}-r^{(2)}_{Cij} = \epsilon^{(2)}_{Tij} - \epsilon^{(2)}_{Cij}$. Note that the terms $\epsilon^{(2)}_{Tij}$ and $\epsilon^{(2)}_{Cij}$ are not assumed to have mean zero. Define $\bar{\tau}_i = (\tau^{(2)}_{i1}+\tau^{(2)}_{i2})/2$ and $\bar{\epsilon}^{(2)}_{ij} = (\epsilon^{(2)}_{Tij}+\epsilon^{(2)}_{Cij})/2$.  

The responses from period 1 are left unmodified relative to their form in \S \ref{ssec.add}, as the treatment does not affect responses in the first period. The observed difference-in-differences $D_i$ is now\begin{align*}
D_i &= (Z_{i1}^{(2)} - Z_{i2}^{(2)})(R_{i1}^{(2)}-R_{i2}^{(2)}) - (Z_{i1}^{(1)} - Z_{i2}^{(1)})(R_{i1}^{(1)}-R_{i2}^{(1)})\\
&= \bar{\tau}_i + (Z_{i1}^{(2)} - Z_{i2}^{(2)})(\bar{\epsilon}_{i1}^{(2)}-\bar{\epsilon}_{i2}^{(2)}) - (Z_{i1}^{(1)} - Z_{i2}^{(1)})(\epsilon_{i1}^{(1)}-\epsilon_{i2}^{(1)}).
\end{align*}The model for a sensitivity analysis developed in \S\S 3.1-3.3 is then assumed to hold with $\bar{\epsilon}^{(2)}_{ij}$ in place of $\epsilon^{(2)}_{ij}$, and with the set $\mathcal{F}_\epsilon$ redefined as $\mathcal{F}_\epsilon = \{x_{ij}^{(t)}, u_{ij}^{(t)},\epsilon^{(2)}_{Cij}, \epsilon^{(2)}_{Tij}, \epsilon^{(1)}_{ij}: i=1,..,I; j=1,2; t=1,2\}$. 

Consider the conventional sensitivity analysis based on (\ref{eq:sensconv}) holding at $\Gamma$, which bounds the probabilities of group assignment given $\cF_\epsilon$ in each pair. As before, we condition on $\mathcal{Z}$, the paired design, along with the value for $B_i = b_i$, such that $D_i = \bar{\tau}_i + V_i\{(\bar{\epsilon}_{i1}^{(2)}-\bar{\epsilon}_{i2}^{(2)}) - b_i(\epsilon_{i1}^{(1)}-\epsilon_{i2}^{(1)})\}.$ We consider a sensitivity analysis for the composite null hypothesis
\begin{align*}
H_0: \bar{\tau} = \frac{1}{I}\sum_{i=1}^I\bar{\tau}_i = \tau_0,
\end{align*} i.e. that the sample average of the treatment effects equals $\tau_0$.  Unlike under Fisher's sharp null, the value for $(\bar{\epsilon}_{i1}^{(2)}-\bar{\epsilon}_{i2}^{(2)}) - b_i(\epsilon_{i1}^{(1)}-\epsilon_{i2}^{(1)})$ is not imputed under the null hypothesis, such that the bounding reference distribution for the average of the differences-in-differences $\bar{D}$ cannot be directly constructed under $H_0$. Nonetheless, a modification of the argument in \citet{fogary2017avg} demonstrates that a valid sensitivity analysis for $\bar{\tau}$ when (\ref{eq:sensconv}) is assumed to hold at $\Gamma$ may be attained by applying Algorithm 2 of \citet{fogary2017avg} at $\Gamma^2$, with the treated-minus-control paired differences (referred to as $\hat{\tau}_i$ in the description of Algorithm 2) replaced with the adjusted differences-in-differences $D_i - \tau_0$. Theorems 1-3 in that work then apply, such that the resulting procedure applied to the differences-in-differences at $\Gamma^2$ provides an asymptotically valid sensitivity analysis for the sample average treatment effect when (\ref{eq:sensconv}) holds at $\Gamma$. We demonstrate how these methods can be applied in one of the applications that follow. While sample average treatment effects allow for arbitrary treatment effect patterns, averages have well known disadvantages when distributions are skewed or have heavy tails. As such, we also illustrate how test statistics based upon quantities other than averages can also be useful in detecting departures from the sharp null.

\subsection{Binary Outcomes}
\label{ssec.bin}

Next, we consider the case when the outcome variable is binary as it is in the election day registration application. In general, DID estimation and inference for discrete outcomes is more complex due to the fact that identification under the DID device is functional form dependent. For example, researchers have shown that standard nonlinear models such as GLMs are not consistent with the common trend assumption necessary for identification under the DID device \citep{puhani2012treatment,lechner2011estimation}. While DID methods for discrete outcomes have been developed \citep{blundell2009alternative,Athey:2006}, \citet{lechner2011estimation} notes that they generally see little use due to their complexity. From our reading of the literature, most analysts apply linear probability models (LPMs).  However, the inferential properties of the LPM will not always be valid when applied to DID \citep{Donald:2007,Bertrand:2004}. Next, we develop a test under the sharp null for the DID device with binary outcomes that is consistent with both a model for additive bias and our matching scheme. 

With binary outcomes, the additive model of bias previously invoked above is no longer appropriate. Nonetheless, inference akin to that proposed in the case of continuous outcomes can be justified when the outcomes are binary through a similar model. Consider the following model for $\P(r_{Cij}^{(t)}=1\mid Z_{ij}^{(t)},\cC)$
\begin{align*}
\P(r_{Cij}^{(t)}=1\mid Z_{ij}^{(t)}, \cC)&= \frac{\exp\{\mu_i + \beta_iZ_{ij} + \alpha_i\1\{t=2\} + \delta_t u_{ij}^{(t)}\}}{1+\exp\{\mu_i + \beta_iZ_{ij} + \alpha_i\1\{t=2\} + \delta_t u_{ij}^{(t)}\}}
\end{align*}

In any quadruplet, the probability of the vector $\mathbf{r}_{Ci} = (r_{Ci1}^{(1)}, r_{Ci2}^{(1)}, r_{Ci1}^{(2)}, r_{Ci2}^{(2)})^T$ is
\begin{align*}
\P(\br_{Ci}\mid \bZ_i, \cC) &= \frac{\exp\{\beta_i\bZ^T_i\br_{Ci} + \mu_i(r_{Ci1}^{(1)}+r_{Ci2}^{(1)}) + (\mu_i+\alpha_i)(r_{Ci1}^{(2)}+r_{Ci2}^{(2)})+\sum_{t=1}^2\delta_t\sum_{j=1}^2u_{ij}^{(t)}r_{Cij}^{(t)}\}}{1+\exp\{\beta_i\bZ^T_i\br_{Ci} + \mu_i(r_{Ci1}^{(1)}+r_{Ci2}^{(1)}) + (\mu_i+\alpha_i)(r_{Ci1}^{(2)}+r_{Ci2}^{(2)}) + \sum_{t=1}^2\delta_t\sum_{j=1}^2u_{ij}^{(t)}r_{Cij}^{(t)}\}}
\end{align*}

Observe that by conditioning on (i) $r_{Ci1}^{(1)}+r_{Ci2}^{(1)}$; (ii) $r_{Ci1}^{(2)}+r_{Ci2}^{(2)}$; and (iii) $\bZ_i^T\br_{Ci}$, we remove dependence on the nuisance parameters $\alpha_i$, $\beta_i$, and $\mu_i$, leaving the resulting conditional distribution dependent solely upon $\{\delta_tu_{ij}^{(t)}: j=1,2; t=1,2\}$. If $\delta_1=\delta_2=0$, the conditional distribution would be entirely free of unknown parameters. As such, this model contains the case of purely additive bias on the logit scale as a special case.

Next, we collect (i), (ii), and (iii) into the set $\mathcal{R}$. Conditioning upon $\cR$, we observe
\begin{align}\label{eq:indep}
\P(\br_{Ci}\mid \bZ_i, \cC, \cR) &= \frac{\exp\{ \sum_{t=1}^2\delta_t\sum_{j=1}^2u_{ij}^{(t)}r_{Cij}^{(t)}\}}{\sum_{\bq \in \cR} \exp\{\sum_{t=1}^2\delta_t\sum_{j=1}^2u_{ij}^{(t)}q_{ij}^{(t)}\} },
\end{align}
such that $\br_{Ci} \indep \mathbf{Z}_i \mid \cC, \cR$. Returning attention to the matched structure by additionally conditioning upon $\cZ$ implies that $\sum_{t=1}^2\sum_{j=1}^2Z_{ij}^{(t)} = 2$ for all $i$. In light of this, the set $\cR$ contains a singleton given $\cZ$ unless the following three conditions hold:
\begin{enumerate}
\item $r_{Ci1}^{(1)} + r_{Ci2}^{(1)} = 1$
\item $r_{Ci1}^{(2)} + r_{Ci2}^{(2)} = 1$
\item $\bZ_i^T\br_{Ci}=1$
\end{enumerate}

The first two conditions are familiar based on the behavior of McNemar's test: only discordant pairs contribute to the null distribution. The third condition requires that not only are the before and after pairs both discordant, but that they are discordant in different ways: one pair in the quadruplet needs to have only the treated unit have the event, while the other needs to have only the control unit have the event. It is only within these pairs where evidence for a treatment effect can be disentangled from the nuisance parameters $\mu_i$, $\alpha_i$, and $\beta_i$.

Next, we consider the difference-in-differences contrast
\begin{align*}
D_i = (Z_{i1}^{(2)} - Z_{i2}^{(2)})(r_{Ci1}^{(2)}-r_{Ci2}^{(2)}) - (Z_{i1}^{(1)} - Z_{i2}^{(1)})(r_{Ci1}^{(1)}-r_{Ci2}^{(1)}).
\end{align*}
If we condition on $\cR$ and $\cZ$, $D_i$ can only take on values $\pm 2$. Further, we observe:
\begin{align}\label{eq:bdelta1}
&\P\left\{r_{Ci1}^{(2)} - r_{Ci2}^{(2)} - (r_{Ci1}^{(1)} - r_{Ci2}^{(1)}) = 2\mid \cC, \cR, \mathcal{Z}\right\} \\\nonumber&= \frac{\exp\{\delta_2(u^{(2)}_{i1}-u^{(2)}_{i2})\}}{\exp\{\delta_1(u^{(1)}_{i1}-u^{(1)}_{i2})\}}\P\left\{r_{Ci1}^{(2)} - r_{Ci2}^{(2)} - (r_{Ci1}^{(1)} - r_{Ci2}^{(1)}) =-2\mid \cC, \cR, \cZ\right\}\end{align}
and
\begin{align}\label{eq:bdelta2}
&\P\left\{r_{Ci1}^{(2)} - r_{Ci2}^{(2)} - (r_{Ci2}^{(1)} - r_{Ci1}^{(1)}) = 2\mid \cC, \cR,\mathcal{Z}\right\}\\\nonumber &= \frac{\exp\{\delta_2(u^{(2)}_{i1}-u^{(2)}_{i2})\}}{\exp\{\delta_1(u^{(1)}_{i2}-u^{(1)}_{i1})\}}\P\left\{r_{Ci1}^{(2)} - r_{Ci2}^{(2)} - (r_{Ci2}^{(1)} - r_{Ci1}^{(1)}) =-2\mid \cC, \cR, \cZ\right\}
\end{align}
\noindent As before, $\delta_1=\delta_2=0$ corresponds to the case of symmetry or the case where the additive model holds, and the DID device successfully removes any hidden bias.

We again use (\ref{eq.trt}) to characterize the unknown probability that a subject is exposed to treatment. As before, we let $V_i = (Z^{(2)}_{i1}-Z^{(2)}_{i2})$ and $B_i = (Z^{(2)}_{i1}-Z^{(2)}_{i2})(Z^{(1)}_{i1}-Z^{(1)}_{i2})$, and we condition on $\bB$. The observed DID contrast is again
\begin{align*}
D_i &= V_i\left(r_{Ci1}^{(2)}-r_{Ci2}^{(2)} - b_i(r_{Ci1}^{(1)}-r_{Ci2}^{(1)})\right)
\end{align*} Suppose that $1/\Lambda \leq \exp\{\lambda_1\}, \exp\{\lambda_2\} \leq \Lambda$, and that $1/\Delta \leq \exp\{\delta_1\}, \exp\{\delta_2\} \leq \Delta$. Analogous derivations to those in \S \ref{ssec.twoparam} result in the bounds
\begin{align}\label{eq:binarybounds}
\frac{\Delta^2+\Lambda^2}{(1+\Lambda^2)(1+\Delta^2)}\leq \P(S_iV_i = 1 \mid  \cC, \cR, \cZ, \bB) \leq \frac{(\Lambda\Delta)^2+1}{(1+\Lambda^2)(1+\Delta^2)},
\end{align} where $S_i = \left(r_{Ci1}^{(2)}-r_{Ci2}^{(2)} - b_i(r_{Ci1}^{(1)}-r_{Ci2}^{(1)})\right)/2 = \text{sign}\left(r_{Ci1}^{(2)}-r_{Ci2}^{(2)} - b_i(r_{Ci1}^{(1)}-r_{Ci2}^{(1)})\right)$.

In the appendix, we formally prove (\ref{eq:binarybounds}). We also show that a sensitivity analysis can be conducted using a conventional sensitivity analysis for McNemar's test with $\Gamma^{2}$ replacing $\Gamma$. In addition, we contrast our approach with an alternative method developed by \citet{zhang2011using}. 

The sensitivity analysis for DID with binary outcomes is straightforward to implement using McNemar's test. If we re-order the quadruplets such that the first $J$ of them satisfy conditions (i)-(iii), then $\sum_{i=1}^J(S_iV_i+1)/2$ is McNemar's statistic. In practical terms, the sensitivity analysis is implemented by applying McNemar's test to matched quadruplets where (a) both pairs are discordant and (b) one pair has treated unit with the outcome, but the other has the control receiving the outcome. Due to the inherent limitations of the model of additive biases with binary outcomes, however, our method does not readily extend to a test for the sample average treatment effect in this case. We leave the development of such a method as an avenue for future research.



\subsection{Remarks on Interpretation}
\label{ssec.rem}

One might interpret the fact that one can use $\Gamma^{2}$ applied to a paired test statistic to find bounds at $\Gamma$ for the DID device to mean that an analysis based on the DID device is more sensitive to hidden bias. For example, imagine that an investigator decided to not employ the DID device and applied a conventional sensitivity analysis for a paired design using data from the posttreatment time period. This conventional sensitivity analysis would start from the premise that $Z_{ij}^{(t)}\indep r^{(t)}_{Cij}\mid \mathcal{C}$, and would model departures from strong ignorability through a model on $\P(Z_{ij}^{(t)}=1 \mid \cF)$, where $\cF = \{x_{ij}^{(t)}, u_{ij}^{(t)}, r^{(t)}_{Cij}\}$. In contrast, the sensitivity analysis for DID conditions on $\cF_\epsilon = \{x_{ij}^{(t)}, u_{ij}^{(t)}, \epsilon_{ij}^{(t)}\}$, that is the DID sensitivity analysis conditions on the residuals instead of the potential outcomes. Since each form of sensitivity analysis has a different conditioning set, the $\Gamma$ from a conventional paired model and the $\Gamma$ from the difference-in-difference model are not directly comparable. That is, $\Gamma=1$ corresponds to a different set of assumptions for each design. For the DID device, $\Gamma=1$ still allows for dependence between $r_{Cij}^{(t)}$ and $Z_{ij}^{(t)}$ given $x_{ij}^{(t)}$, while for the paired sensitivity analysis these would be assumed conditionally independent at $\Gamma=1$. As a result, the minimal $\Gamma$'s for which the models hold might be considerably different.

More concretely, suppose we conducted a sensitivity analysis assuming only a paired structure, and found the changepoint $\Gamma$ to be $2.25$. Then using the DID device, we found the changepoint to be at $\Gamma^2=2.25$, corresponding to $\Gamma=1.5$ for the DID analysis. Should investigators interpret a result of this type as evidence that a paired design is superior to using the DID device? The answer is only an unequivocal yes if the DID device was unnecessary and does not remove any bias. While it is implausible that DID removes all bias, it is also implausible that use of DID doesn't remove any bias. Moreover, knowing whether DID is unnecessary is also not a testable proposition. As such, we caution against interpreting our results as a clear indication that DID is subject to greater bias than a conventional matched design.


\section{Application: Disability Payments in Germany}
\label{sec.germ}

For the first study, we re-analyzed the data from \citet{puhani2010}. As we outlined above, their study applied DID to understand whether less generous disability payments caused to workers miss fewer work days. We begin the analysis with plots of the outcomes for the treated and control groups in both the before and after period. Simple plots of this type can be useful to assess whether it appears the temporal path of the treated group appears to deviate from a common trend. While we observe a clear decline in the number of days absent for the treated, we also observe an over time change in the control group outcomes in the \emph{opposite} direction. This pattern does not suggest that treated and control groups follow a common trend.

\begin{figure}[htbp]
\centering
  \subfloat[Raw Outcomes]{\includegraphics[width=0.28\textheight]{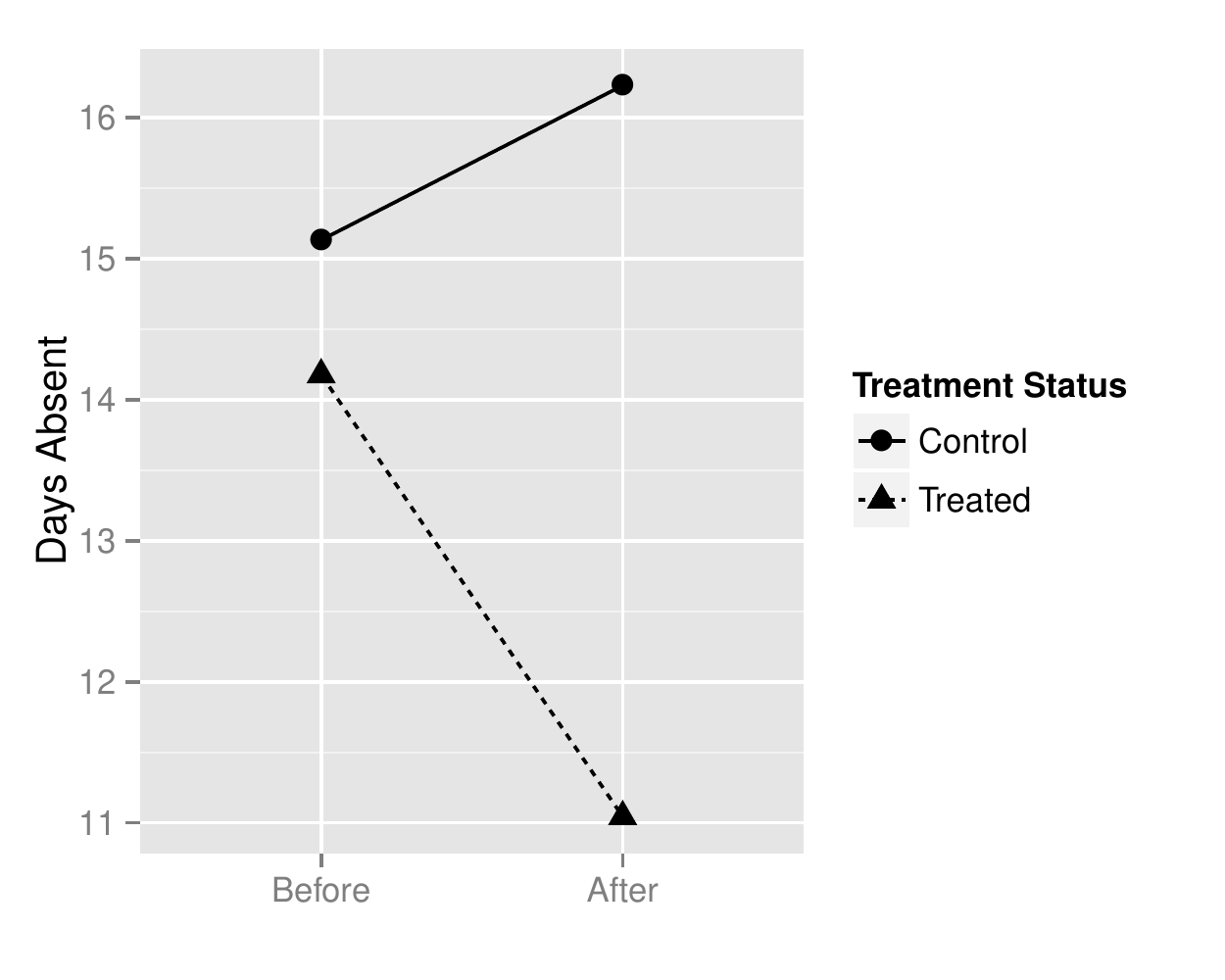}\label{fig:soepa}}
  \subfloat[Log Scale]{\includegraphics[width=0.28\textheight]{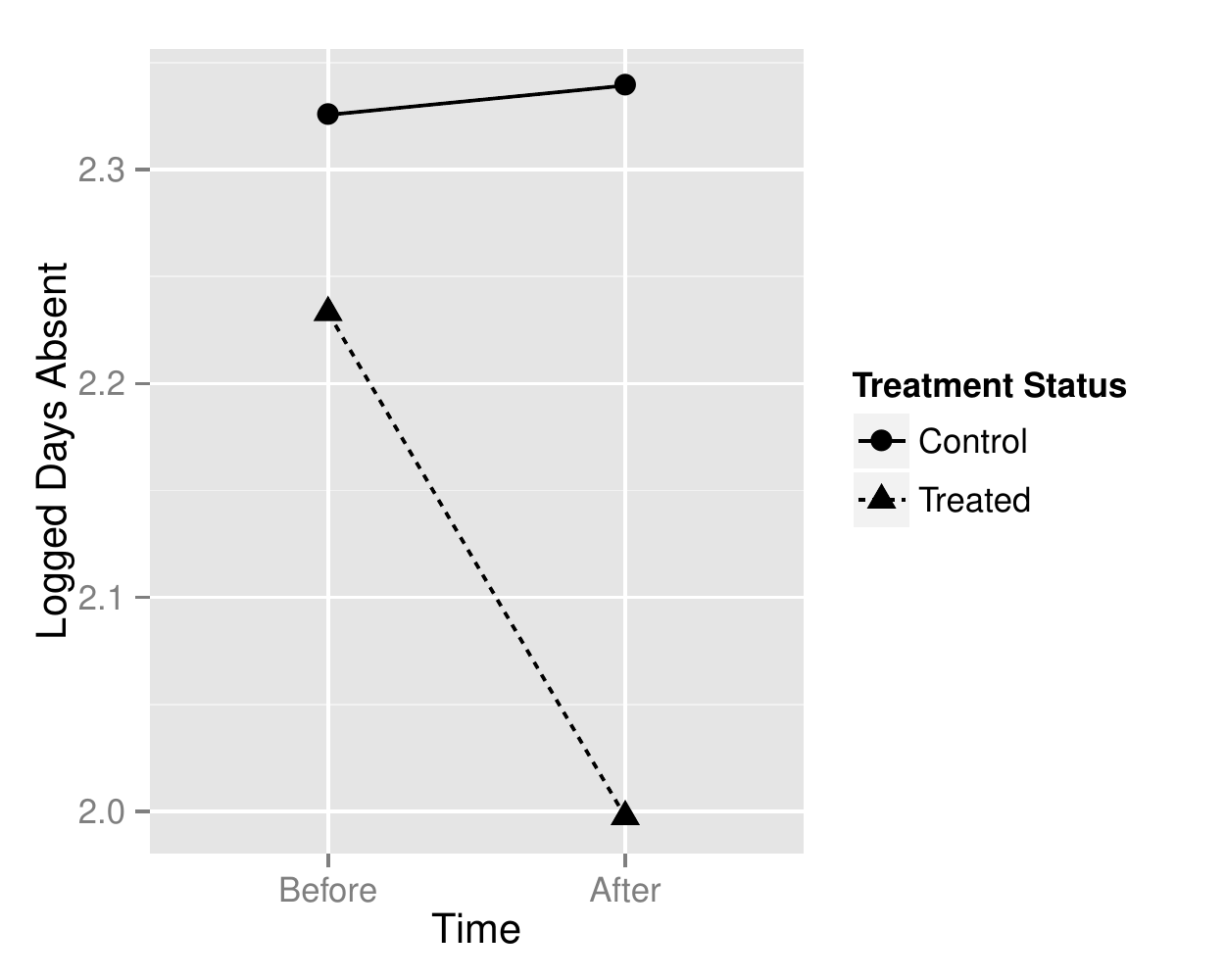}\label{fig:soepb}}
\caption{Outcomes for the German Disability Payments}
\label{fig:cond.plots}
\end{figure}

To adjust for observed covariates, we implemented the matching plan outlined in \S \ref{ssec.match}. We match on the same set of covariates used in the original analysis. These include measures for hourly wages, age, education levels, blue or white collar status, firm size, length of tenure with company, and industry. For three nominal covariates, we set fine balance constraints. In our match, we applied near fine balance constraints on firm size and length of tenure with one's employer; we finely balanced industry. The matching resulted in 356 matched pairs in the period before treatment went into effect. We implemented the match via cardinality matching, which returns the largest set of matched pairs that met our pre-specified balance constraints \citep{Zubizarreta:2014}.

Next, we applied the exact same form of matching to the treated and control units in the period after the change in disability payments. The match in the post-treatment period produced 474 matched pairs.  Finally, we matched the 354 pairs from the pretreatment period to the 474 matched pairs from the post-treatment period. To match pairs to pairs, we took within pair averages within sets of matched pairs. In this match, we exactly matched on industry and finely balanced both firm size and length of tenure with company.  This resulted in 331 sets of pairs matched to pairs. Balance tables from all matches are in the appendix.  

The estimated DID treatment effect based on the averages of the differences-in-differences is -2.3, which would imply that reducing disability payments reduced the average number of days absent from work by just over two days should the assumptions underpinning differences-in-differences hold. That said, we are unable to reject the weak null that the sample average treatment effect is zero while employing the method of \ref{ssec.het} at $\Gamma=1$ ($p = .132$). The distribution for the number of days absent has a long tail, and averages (employed in the aforementioned test) have known deficiencies in the presence of long tails. Hence, we also use the signed rank test to test the sharp null hypothesis of no treatment effect at all. We are also unable to reject the sharp null hypothesis with $p=0.057$. Assuming constant effects using the signed rank test, the Hodges-Lehmann point estimate is -2.0, with a 95\% confidence interval of $[-4.5, 0.50]$.  In sum, while we cannot reject the sharp null hypothesis through either method, the analyses based on raw differences-in-differences and based upon a signed rank test both estimate that a reduction in disability payments reduced the number of days absent among the treatment group by two days.

The above estimates assume that only additive hidden bias is present (i.e. that the parallel trends assumption holds). To assess robustness of both hypothesis tests and point estimates to this assumption, we conduct a sensitivity analysis for the DID treatment effect estimate. As we outlined in \S \ref{sec.sens}, we can apply standard methods for Rosenbaum bounds using $\Gamma^2$ to calculate sensitivity at $\Gamma$. For a sensitivity analysis for the sample average treatment effect, the upper-bound on the one-sided $p$-value at $\Gamma=1.01$ is 0.075. Here, the bounds for the p-value are less useful given we are unable to reject the sharp null when $\alpha = 0.05$ even at $\Gamma=1$. Nonetheless, as the point estimate would reflect a large effect we may still be interested in the range of possible point estimates once hidden bias is allowed to corrupt the inference. We find that the bounds on the point estimate on are 0 and 4.1 for $\Gamma = 1.11$. Thus, the estimate is extremely sensitive to bias from a hidden confounder. A hidden confounder would have to change the odds of treatment within matched pairs of pairs by only a small amount to alter our conclusions.

\section{Application: Election Day Registration}
\label{sec.edr}

In studies of election administration, a number of studies have concluded that EDR has contributed to an increase in voter turnout \citep{Brians:1999, Brians:2001,Hanmer:2007,Hanmer:2009,Highton:1998, Knack:2001, Mitchell:1995, Rhine:1995, Teixeira:1992, Timpone:1998, Wolfinger:1980}. However, recent works suggest these studies are subject to substantial bias from hidden confounders \citep{Keele:2010}. As an illustration, we conduct a small scale study of EDR. In our study, we focus on Wisconsin, one of the first states to adopt EDR, and where the effect of EDR is widely understood to have contributed to an increase in turnout \citep{Hanmer:2009}. 

The data are extracts from the 1972 and 1980 Current Population Survey (CPS) and are a subset of the data from \citet{Keele:2010}. The CPS is a monthly individual level survey conducted by the U.S. census which asks respondents about voting in the November survey of election years. Wisconsin first used EDR in 1976, and we use turnout levels in the 1980 presidential election as the post-treatment period in case of any delay in the effect of EDR. We use voters from Illinois as controls. Illinois would seem to be a reasonable counterfactual for Wisconsin: it is adjacent to Wisconsin and both states have large metropolitan areas with minority communities, but both have large rural populations as well.

\begin{figure}[htbp]
\centering
\includegraphics[scale = .6]{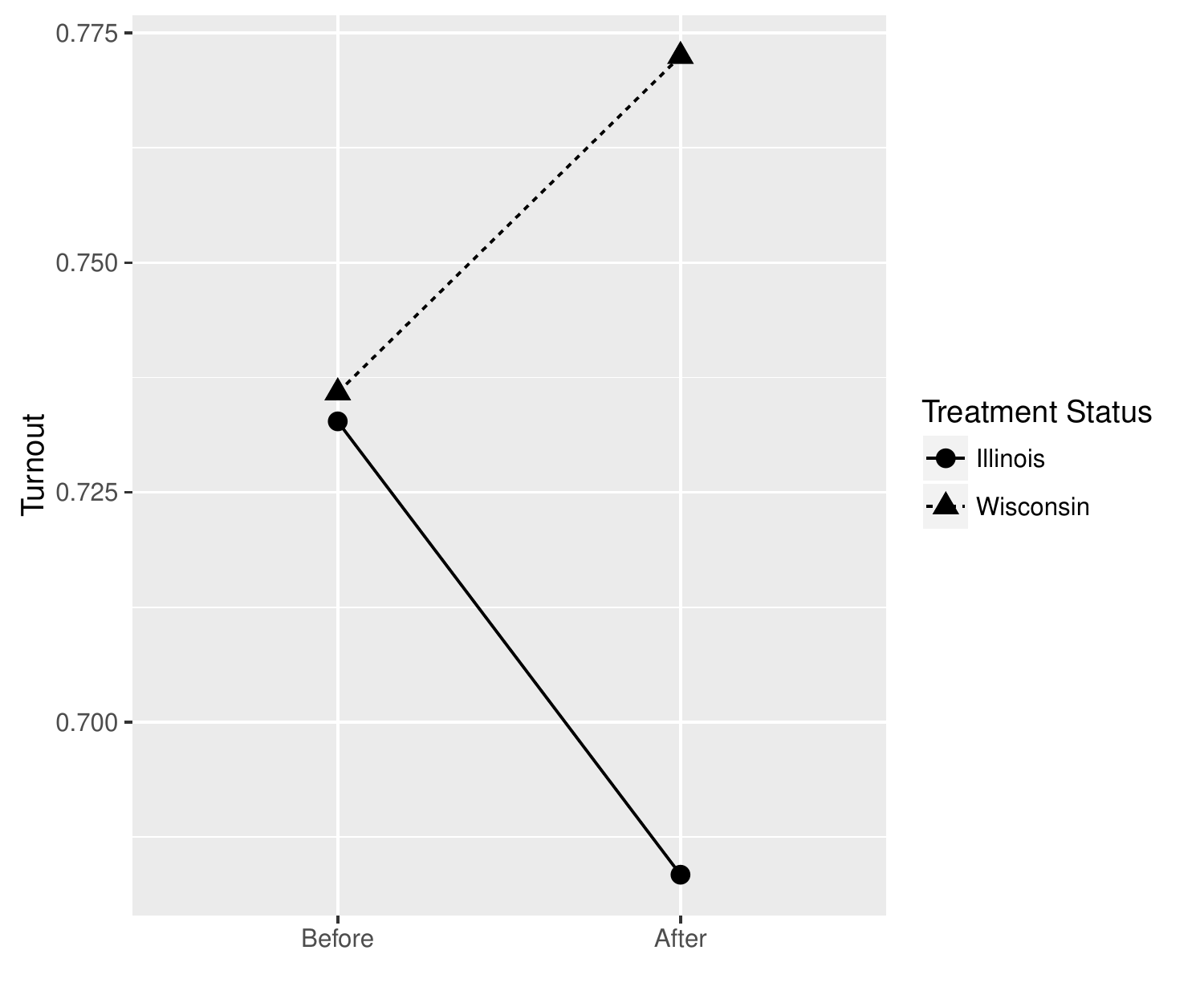}
\caption{Outcomes for the EDR Example. Before Year is 1972, After Year is 1980.}
\label{fig:edr.plots}
\end{figure}

As before, we begin with a plot of the turnout rates in both states. Figure~\ref{fig:edr.plots} contains the turnout before and after the implementation of EDR in Wisconsin for both states. First, we observe a sharp increase in turnout in Wisconsin in 1980, which suggests that perhaps EDR did increase turnout in the state. However, the plot suggests that some other factor or factors contributes to a sharp decrease in turnout in Illinois between 1972 and 1980. This pattern might be a result of voter mobilization efforts in Wisconsin in 1980 and in Illinois in 1972. A bias of this form  would tilt higher responses in Wisconsin 1980 and Illinois in 1972 and lower in Illinois in 1980 and Wisconsin in 1972.

We begin the analysis by matching Wisconsin residents to Illinois residents, first in 1972, and then again in 1980.  We match residents on age, an indicator if he or she is African American, female, a categorical scale of education, a categorical scale of income, and an interaction between education and income categories. In our match, we matched exactly on whether a resident was African-American, and we applied near--fine balance to education, income, and the interaction between education and income categories.  We allowed for a deviation of two categories on the near--fine balance in the match. After matching in 1972, we have 1427 matched pairs.  After matching in 1980, we have 1718 matched pairs.  We then performed the pair-to-pair match, where we matched the pairs from 1972 to pairs from 1980. Imbalances tended to be much larger across the two time periods than within each year. For the pair-to-pair match, we again applied cardinality matching. We are left with 938 matched pairs from 1972 matched to 938 matched pairs from 1980. 

First, we calculated the DID treatment effect by applying the usual DID contrast to the outcomes for the matched quadruples. According to this estimate, the turnout rate increased 10.3 percentage points in Wisconsin as compared to Illinois, and we can reject the null ($p <.001$). Next, we calculate McNemar's statistic to test the sharp null as described in \S \ref{ssec.bin}. Under the test for the sharp null, we easily reject ($p = 0.018$). What might account for the difference between the conventional test and the test of the sharp null? Most likely the difference stems from the fact that the conventional test is known to underestimate statistical uncertainty \citep{Donald:2007}.

Both estimates assume that there is no bias from hidden confounders beyond those which the parallel trends assumption entitle us to ignore. To assess the robustness of our findings to this assumption, we conduct a sensitivity analysis. The sensitivity bound is calculated with McNemar's test using $\Gamma^2$ to find the changepoint value for $\Gamma$. We find that the one-sided $p$-value is 0.051 when $\Gamma \approx 1.05$. This implies that an unobserved confounder could reverse our conclusions if it affected the odds of assignment to treatment or control in either time period by 5\%. The EDR application clearly illustrates the importance of a sensitivity analysis when investigators use the DID device. The conventional analysis appears to make a compelling case for EDR increasing turnout in Wisconsin; however, this estimate rests on the assumption of parallel trends. The sensitivity analysis reveals that a deviation from the parallel trends assumption could easily explain these results, and thus serves as important check on the plausibility of a potential causal finding.

\section{Discussion}

The method of DID is widely used to estimate causal effects. The two applications we present are emblematic of areas where DID is used. The first is a change in labor laws in Germany, and the second is a change in election laws in the United States. Under this device, the hope is that the configuration of the bias from unobserved confounders has a specific additive form that can be eliminated when the investigator obtains data from treated and control groups before and after a treatment goes into effect. Here, we outlined how covariate adjustment based on matching makes much weaker functional form assumptions than the usual methods based on regression models. Next, we outlined a method of sensitivity analysis. Importantly, the sensitivity analysis is easy to implement using existing methods and software, and reveals how an unobserved confounder can change the odds of treatment assignment through two different paths.
    
Finally, we think it is worth emphasizing that there is typically nothing haphazard or as-if random about treatment assignment in most applications that use DID. It is for this reason that we refer to DID as a device and not a type of natural experiment. The plausibility of designs that employ the DID device should be judged based on the assignment process and how well it can be modeled rather than whether or not it possible to use the DID device. In both of the applications analyzed here, policy-makers made these changes for reasons that are far from random or haphazard. A useful contrast is between the DID device and the regression discontinuity (RD) design. In an RD design, a known treatment assignment rule is applied and respected. The strength of RD designs comes directly from the use and application of this known assignment rule \citep{Lee:2010}. Under DID, the treatment assignment rule is typically far more ambiguous leading to far more ambiguous conclusions. 

\clearpage

\singlespacing
\bibliographystyle{\string~/Dropbox/texmf/bibtex/bst/asa}
\bibliography{\string~/Dropbox/texmf/bibtex/bib/keele_revised2}

\clearpage

\appendix
\appendixpage
\addappheadtotoc

\def\thesection{A.\arabic{section}}

\section{Further Results for Binary Outcomes}

Here we develop further results for the case with binary outcomes.  First, we formally derive the two-parameter sensitivity analysis with binary outcomes. Then, we describe how a conventional sensitivity analysis can be used to derive bounds at $\Gamma$ using $\Gamma^2$.  Finally, we formally compare our method with results from \citet{zhang2011using}.

\subsection{Deriving the Two-Parameter Sensitivity Analaysis}

Note $A_i = |r_{Ci1}^{(2)}-r_{Ci2}^{(2)} - b_i(r_{Ci1}^{(1)}-r_{Ci2}^{(1)})|$ is fixed at 2 whenever $\bR$ is not a singleton given $\cC$. Let $S_i = \text{sign}\left(r_{Ci1}^{(2)}-r_{Ci2}^{(2)} - b_i(r_{Ci1}^{(1)}-r_{Ci2}^{(1)})\right)$. Note by (\ref{eq:indep}) that $\br_{Ci}\indep \bZ_i \mid \cC, \cR$, such that $S_i \indep V_i \mid \cC, \cR, \cZ, \bB$. The derivations for sensitivity bounds with on continuous outcomes will now carry through. Let $\eta_{is} = \P(S_i = s\mid \cC, \cZ, \cR, \bB)$ and consider the case where $\cR\mid \cZ$ is not a singleton:
\begin{align*}
\eta_{1i} &= \frac{\exp\{\delta_2(u^{(2)}_{i1}-u^{(2)}_{i2}) - b_i\delta_1(u^{(1)}_{i1}-u^{(1)}_{i2})\}}{1+\exp\{\delta_2(u^{(2)}_{i1}-u^{(2)}_{i2}) - b_i\delta_1(u^{(1)}_{i1}-u^{(1)}_{i2})\}}\\
\eta_{-1i} &= 1-\eta_{1i}.
\end{align*}
Further let $\rho_i = \P(V_i=1 \mid \cC, \cZ, \cR, \bB$), which we write as
\begin{align*}
\rho_i &= \frac{\exp\{\lambda_2(u^{(2)}_{i1}-u^{(2)}_{i2}) + b_i\lambda_1(u^{(1)}_{i1}-u^{(1)}_{i2})\}}{1+\exp\{\lambda_2(u^{(2)}_{i1}-u^{(2)}_{i2}) + b_i\lambda_1(u^{(1)}_{i1}-u^{(1)}_{i2})\}}.\\
\end{align*} 
\noindent Finally, with these terms we write:
\begin{align}\label{eq:bprob}
&\P(S_iV_i = 1 \mid \cC, \cZ, \cR, \bB)\nonumber\\
&= \rho_i\eta_{1i} + (1-\rho_i)\eta_{-1i}\\\nonumber
&= \frac{\exp\{(\lambda_2+\delta_2)(u^{(2)}_{i1}-u^{(2)}_{i2}) + b_i(\lambda_1-\delta_1)(u^{(1)}_{i1}-u^{(1)}_{i2})\}+1}{[1+\exp\{\lambda_2(u^{(2)}_{i1}-u^{(2)}_{i2}) + b_i\lambda_1(u^{(1)}_{i1}-u^{(1)}_{i2})\}][1+\exp\{\delta_2(u^{(2)}_{i1}-u^{(2)}_{i2}) - b_i\delta_1(u^{(1)}_{i1}-u^{(1)}_{i2})\}]}.
\end{align}
Suppose $1/\Lambda \leq \exp\{\lambda_1\}, \exp\{\lambda_2\} \leq \Lambda$, and that $1/\Delta \leq \exp\{\delta_1\}, \exp\{\delta_2\} \leq \Delta$. It is straightforward to show that given $u_{ij}^{(t)}\in [0,1]$, 
\begin{align*}
\frac{\Delta^2+\Lambda^2}{(1+\Lambda^2)(1+\Delta^2)}\leq \P(S_iV_i = 1 \mid  \cC, \cZ, \cR, \bB) \leq \frac{(\Lambda\Delta)^2+1}{(1+\Lambda^2)(1+\Delta^2)}
\end{align*}
These worst-case bounds are then consistent with both binary outcomes and the model of additive bias assumed under the DID device.

\subsection{A Conventional Sensitivity Analysis}

Suppose we start from a model on $\P(Z_{ij}^{(t)} = 1 \mid \cF)$, where $\cF = \{x_{ij}^{(t)}, u_{ij}^{(t)}, r_{Cij}^{(t)}\}$.
\begin{align} \label{eq:sensbinaryconv}
\P(Z_{ij}^{(t)} = 1 \mid \cF) &= \frac{\exp\{\kappa_t(x^{(t)}_{ij}) + \varrho_i r_{Cij} + \gamma_t u_{ij}^{(t)}\}}{1+\exp\{\kappa_t(x_{ij}) + \varrho_ir_{Cij} + \gamma_t u_{ij}^{(t)}\}}
\end{align}

From this, we have the probability of the quadruple-specific treatment vector 
\begin{align*}
\P(\bZ_i=\bz_i\mid \cF) &= \frac{\exp\{\varrho_i\bz^T_i\br_{Ci} + \sum_{t=1}^2\kappa_t(x^{(t)}_{ij})(z_{i1}^{(t)}+z_{i2}^{(t)}) +\sum_{t=1}^2\gamma_t\sum_{j=1}^2u_{ij}^{(t)}z_{ij}^{(t)}\}}{1+\exp\{\varrho_i\bz^T_i\br_{Ci} + \sum_{t=1}^2\kappa_t(x^{(t)}_{ij})(z_{i1}^{(t)}+z_{i2}^{(t)}) + \sum_{t=1}^2\gamma_t\sum_{j=1}^2u_{ij}^{(t)}z_{ij}^{(t)}\}}
\end{align*}

We see that by conditioning on $Z^{(1)}_{i1}+Z^{(1)}_{i2}$, $Z^{(2)}_{i1}+Z^{(2)}_{i2}$, and $\bZ_i^T\br_{Ci}$, we remove dependence on all nuisance parameters spare $\{\gamma_t u_{ij}^{(t)}\}$.

The first two conditions are standard for the paired design (we condition on $Z_{i1}+Z_{i2}=1$). Once again, we assume the following condition holds: $\bZ_i^T\br_{Ci}=1$. Recall that $\br_{Ci}$ are considered fixed through conditioning on $\cF$. So, this implies that inference is restricted to quadruplets where (a) both pairs are discordant and (b) one pair has treated unit with the outcome, but the other has the control receiving the outcome. To see why, first consider $\bZ^T\br_{Ci} = 0$. In this case, both treated individuals across the quadruplets did not have the event. So, even if the pairs themselves were discordant, conditioning on $\bZ^T\br_{Ci}=0$ means that the controls always have the event. The analogous condition holds for $\bZ^T\br_{Ci}=2$: the treated always have the events, even in discordant pairs. Now, if $\bZ_i^T\br_{Ci}=1$ and only one pair is discordant, this still fixes the assignments. For instance, if the concordant pair both have the outcome, this means in the discordant pair the treated individual has to receive the control. Hence, variation only occurs if $\bZ_i^T\br_{C_i} = 1$ and both pairs are discordant.

Within those quadruplets, we have, in effect, conditioned upon $B_i = b_i$. That is, we only consider assignments where the treated individuals in the quadruplet move together: either both stay treated, or both become controls. As both pairs are discordant, any other assignment would violate the condition $\bZ_i^T\br_{Ci}=1$. So, in these pairs,
\begin{align*}
\P(\bZ_i = \bz_i\mid  \cF, \cZ, \bZ_i^T\br_i=1, \bB) &=
\frac{\exp\{\gamma_1(z_{i1}^{(1)} - z_{i2}^{(1)})(u_{i1}^{(1)}-u_{i2}^{(1)}) + \gamma_2b_i(z_{i1}^{(2)}-z_{i2}^{(2)})(u_{i1}^{(2)}-u_{i2}^{(2)})\}}{1+ \exp\{\gamma_1(z_{i1}^{(1)} - z_{i2}^{(1)})(u_{i1}^{(1)}-u_{i2}^{(1)} )+ \gamma_2b_i(z_{i1}^{(2)}-z_{i2}^{(2)})(u_{i1}^{(2)}-u_{i2}^{(2)})\}}
\end{align*}

Suppose $1/\Gamma \leq \exp(\gamma_1), \exp(\gamma_2) \leq \Gamma$. Then, we readily see
\begin{align*} \frac{1}{1+\Gamma^2}\leq\P(\bZ_i = \bz_i\mid \cF, \bZ, \bZ_i^T\br_i=1, \bB)\leq \frac{\Gamma^2}{1+\Gamma^2}\end{align*}
Of course, this gives the bound $1/(1+\Gamma^2)\leq \P(V_i=1\mid \cF, \bZ, \bZ_i^T\br_i=1, \bB)\leq \Gamma^2/(1+\Gamma^2)$. This suggests conducting a sensitivity analysis under model (\ref{eq:sensconv}) using classical methods at $\Gamma^2$ if $\exp(|\gamma_1|), \exp(|\gamma_2|)\leq \Gamma$. Re-order the quadruplets such that the first $J$ of them satisfy the required conditions. In those pairs, 
$\sum_{i=1}^J(D_i/4 + 1/2)$ would be McNemar's statistic, and we can use methods for sensitivity analysis for McNemar's test at $\Gamma^2$.

\subsection{Why is this different from Gart (1969)?}

\citet{zhang2011using} suggest another way to conduct a sensitivity analysis with bianry outcomes in the context of difference-in-differences. Their proposal amounts to an extension of \citet{gart1969exact} to the case of hidden bias. The difference between their test and the one we developed in the main text stem from the fact that the mode of inference we considered is different. \citet{zhang2011using} consider a \textit{single} slope on $\varrho$, rather than a quadruple-specific slope $\varrho_{i}$ in (\ref{eq:sensbinaryconv}). Their model considers
\begin{align} \label{eq:sensbinarygart}
\P(Z_{ij}^{(t)} = 1 \mid \cF) &= \frac{\exp\{\kappa_t(x_{ij}) + \varrho r_{Cij} + \gamma u_{ij}^{(t)}\}}{1+\exp\{\kappa_t(x_{ij}) + \varrho r_{Cij} + \gamma u_{ij}^{(t)}\}}
\end{align}
Under this model, one can remove dependence on the single parameter $\varrho$ by conditioning on the quantity $\bZ^T\br_C = \sum_{i=1}^I\sum_{j=1}^2\sum_{t=1}^2Z_{ij}^{(t)}r_{Cij}^{(t)}$, and it is this conditional distribution which is employed there to get the extended hypergeometric. While one could also remove dependence even in this model on $\varrho$ by conditioning on $(\bZ_1^T\br_{C1},..., \bZ_I^T\br_{CI})^T$ as we did in our results, this is not minimally sufficient and would not be recommended were one willing to assume a constant value for $\varrho$. If one is truly concerned about heterogeneity in $\varrho_i$ across quadruples, then conditioning on $\bZ^T\br_C$ does not remove dependence on the nuisance parameter, rendering the method extending \citet{gart1969exact} inapplicable. That said, the conditioning required when allowing $\varrho_i$ to vary is substantially finer and may exclude a prohibitively large proportion of matched quadruplets. In these cases, the method of \citet{zhang2011using} provides a sensible alternative sensitivity analysis.

\section{Balance Tables for Applications}

Here, we include balance test results for the various matches completed in the two applications.

\begin{table}[ht]
\centering
\caption{Standardized Differences and p-values for Treated to Control Match in the Pretreatment Period for the Disability Payments Application }
\begin{tabular}{lcccc}
 \toprule
 & \multicolumn{2}{c}{Before Matching} &  \multicolumn{2}{c}{After Matching}\\
 
\cmidrule{2-5}
 & Std Dif & P-val & Std Dif & P-val \\ 
  \midrule
Regional Unemp. & 0.19 & 0.00 & 0.02 & 0.84 \\ 
  Hourly Wage & -0.04 & 0.51 & 0.06 & 0.50 \\ 
  Age & -0.10 & 0.09 & 0.00 & 0.95 \\ 
  Married & -0.02 & 0.70 & 0.06 & 0.41 \\ 
  Female & -0.01 & 0.82 & -0.09 & 0.23 \\ 
  Children Under 16 & -0.03 & 0.62 & 0.00 & 1.00 \\ 
  Female \& Child Under 16 & 0.03 & 0.64 & -0.02 & 0.78 \\ 
  Female \& Married & 0.02 & 0.67 & 0.05 & 0.53 \\ 
  Education & -0.00 & 0.96 & -0.07 & 0.35 \\ 
  Temporary Contract & -0.04 & 0.48 & -0.08 & 0.33 \\ 
  Blue Collar & -0.04 & 0.44 & -0.04 & 0.60 \\ 
  White Collar & 0.15 & 0.01 & 0.02 & 0.82 \\ 
  Civil Servant & -0.28 & 0.00 & 0.04 & 0.31 \\ 
  German & 0.04 & 0.52 & -0.02 & 0.76 \\ 
  West German & -0.22 & 0.00 & -0.02 & 0.81 \\ 
  Satisfaction w Health & 0.00 & 0.99 & -0.07 & 0.32 \\ 
  Self-Reported Health Status & 0.07 & 0.21 & 0.10 & 0.16 \\ 
   \bottomrule
\end{tabular}
\end{table}

\begin{table}[ht]
\centering
\caption{Standardized Differences and p-values for Treated to Control Match in the Post-treatment Period for the Disability Payments Application}
\begin{tabular}{lcccc}
 \toprule
 & \multicolumn{2}{c}{Before Matching} &  \multicolumn{2}{c}{After Matching}\\
 
\cmidrule{2-5}
 & Std Dif & P-val & Std Dif & P-val \\ 
  \midrule
Regional Unemp. & 0.12 & 0.02 & -0.01 & 0.91 \\ 
  Hourly Wage & -0.10 & 0.06 & 0.03 & 0.66 \\ 
  Age & 0.01 & 0.86 & 0.01 & 0.86 \\ 
  Married & 0.05 & 0.34 & 0.07 & 0.25 \\ 
  Female & 0.03 & 0.55 & 0.00 & 0.95 \\ 
  Children Under 16 & 0.10 & 0.04 & 0.00 & 1.00 \\ 
  Female \& Child Under 16 & 0.18 & 0.00 & 0.02 & 0.76 \\ 
  Female \& Married & 0.09 & 0.09 & 0.03 & 0.61 \\ 
  Education & -0.00 & 0.99 & 0.01 & 0.92 \\ 
  Temporary Contract & 0.10 & 0.08 & 0.09 & 0.16 \\ 
  Blue Collar & -0.02 & 0.77 & -0.03 & 0.69 \\ 
  White Collar & 0.12 & 0.02 & 0.01 & 0.85 \\ 
  Civil Servant & -0.24 & 0.00 & 0.04 & 0.37 \\ 
  German & -0.08 & 0.12 & -0.02 & 0.73 \\ 
  West German & -0.17 & 0.00 & -0.01 & 0.83 \\ 
  Satisfaction w Health & -0.02 & 0.62 & -0.01 & 0.85 \\ 
  Self-Reported Health Status & 0.02 & 0.64 & 0.03 & 0.59 \\ 
   \bottomrule
\end{tabular}
\end{table}

\begin{table}[ht]
\centering
\caption{Standardized Differences and p-values for Pair-to-Pair Match in the Disability Payments Application}
\begin{tabular}{lcccc}
 \toprule
 & \multicolumn{2}{c}{Before Matching} &  \multicolumn{2}{c}{After Matching}\\
 
\cmidrule{2-5}
 & Std Dif & P-val & Std Dif & P-val \\ 
  \midrule
Regional Unemp. & 0.05 & 0.47 & 0.02 & 0.83 \\ 
  Hourly Wage & -0.28 & 0.00 & -0.10 & 0.17 \\ 
  Age & -0.16 & 0.02 & -0.05 & 0.54 \\ 
  Married & -0.21 & 0.00 & -0.08 & 0.29 \\ 
  Female & 0.06 & 0.36 & 0.06 & 0.47 \\ 
  Children Under 16 & -0.18 & 0.01 & -0.10 & 0.21 \\ 
  Female \& Child Under 16 & -0.12 & 0.08 & -0.07 & 0.34 \\ 
  Female \& Married & -0.11 & 0.11 & -0.07 & 0.40 \\ 
  Education & 0.17 & 0.01 & 0.10 & 0.17 \\ 
  Temporary Contract & 0.19 & 0.01 & 0.10 & 0.19 \\ 
  Blue Collar & 0.09 & 0.19 & 0.05 & 0.50 \\ 
  White Collar & -0.07 & 0.30 & -0.06 & 0.47 \\ 
  Civil Servant & -0.08 & 0.26 & 0.01 & 0.83 \\ 
  German & 0.09 & 0.22 & 0.04 & 0.61 \\ 
  West German & -0.05 & 0.44 & -0.03 & 0.74 \\ 
  Satisfaction w Health & 0.23 & 0.00 & 0.10 & 0.18 \\ 
  Self-Reported Health Status & -0.07 & 0.34 & 0.01 & 0.88 \\ 
   \bottomrule
\end{tabular}
\end{table}

\begin{table}[ht]
\centering
\caption{Standardized Differences and p-values for Treated to Control Match in the Pretreatment Period for the Election Day Registration Application }
\begin{tabular}{lcccc}
 \toprule
 & \multicolumn{2}{c}{Before Matching} &  \multicolumn{2}{c}{After Matching}\\
 
\cmidrule{2-5}
 & Std Dif & P-val & Std Dif & P-val \\ 
  \midrule
Age & 0.00 & 0.98 & -0.05 & 0.18 \\ 
  African-American & -0.31 & 0.00 & 0.04 & 0.13 \\ 
  Female & -0.01 & 0.72 & -0.04 & 0.28 \\ 
  Education & 0.07 & 0.02 & -0.05 & 0.18 \\ 
  Income & 0.02 & 0.52 & -0.05 & 0.19 \\ 
  Education X Income & 0.07 & 0.02 & 0.05 & 0.19 \\ 
\midrule
\end{tabular}
\end{table}

\begin{table}[ht]
\centering
\caption{Standardized Differences and p-values for Treated to Control Match in the Pretreatment Period for the Election Day Registration Application }
\begin{tabular}{lcccc}
 \toprule
 & \multicolumn{2}{c}{Before Matching} &  \multicolumn{2}{c}{After Matching}\\
 
\cmidrule{2-5}
 & Std Dif & P-val & Std Dif & P-val \\ 
  \midrule
e & -0.06 & 0.04 & 0.05 & 0.15 \\ 
  African-American & -0.24 & 0.00 & -0.04 & 0.13 \\ 
  Female & -0.01 & 0.77 & 0.04 & 0.25 \\ 
  Education & 0.17 & 0.00 & -0.05 & 0.23 \\ 
  Income & 0.11 & 0.00 & 0.05 & 0.17 \\ 
  Education X Income & 0.16 & 0.00 & -0.05 & 0.19 \\ 
   \midrule
\end{tabular}
\end{table}

\begin{table}[ht]
\centering
\caption{Standardized Differences and p-values for Treated to Control Match in the Pair-to-Pair Match for the Election Day Registration Application }
\begin{tabular}{lcccc}
 \toprule
 & \multicolumn{2}{c}{Before Matching} &  \multicolumn{2}{c}{After Matching}\\
 
\cmidrule{2-5}
 & Std Dif & P-val & Std Dif & P-val \\ 
  \midrule
Age & 0.18 & 0.00 & -0.05 & 0.29 \\ 
  African-American & -0.05 & 0.15 & -0.07 & 0.12 \\ 
  Female & 0.10 & 0.00 & -0.02 & 0.67 \\ 
  Education & -0.27 & 0.00 & 0.05 & 0.28 \\ 
  Income & -1.27 & 0.00 & -0.05 & 0.15 \\ 
  Education X Income & -1.10 & 0.00 & -0.05 & 0.18 \\
 \midrule
\end{tabular}
\end{table}

\end{document}